\documentclass[a4paper,12pt, leqno]{article}
\usepackage{amsmath}
\usepackage{amssymb, latexsym}
\usepackage{amsfonts}
\usepackage{makeidx}
\usepackage{fancybox}
\usepackage{amssymb}

\usepackage{longtable} 

\usepackage{color}     

\date{}

\newtheorem{lem}{Lemma}[section]
\newtheorem{thm}{Theorem}[section]
\newtheorem{prop}{Proposition}[section]
\newtheorem{cor}{Corollary}[section]
\newtheorem{definition}{Definition}[section]

\numberwithin{equation}{section}
\newcommand{\dbar}{d\!\!\!{\lower-0.6ex\hbox{$-$}}\!}
\newcommand{\dslash}{d\!\!\!{\lower-0.6ex\hbox{$-$}}}
\newcommand{\e}{\varepsilon}
\newcommand{\h}{\hbar}
\newcommand{\ott}{\lower-0.4ex\hbox{${\scriptscriptstyle{\otimes}}$}}
\newcommand{\btt}{\lower-0.2ex\hbox{${\scriptscriptstyle{\bullet}}$}}
\newcommand{\ctt}{\lower-0.2ex\hbox{${\scriptscriptstyle{\circ}}$}}
\newcommand{\dtt}{\lower-0.2ex\hbox{${\scriptscriptstyle{\diamond}}$}}
\newcommand{\odt}{\lower-0.4ex\hbox{${\scriptscriptstyle{\odot}}$}}

\newcommand{\Cal}{\mathcal}
\begin{document}
\pagestyle{plain}

\title{Deformation Expression for Elements of Algebras (I)\\
--(Jacobi's theta functions and $*$-exponential functions)--}
\author{
     Hideki Omori\thanks{ Department of Mathematics,
             Faculty of Sciences and Technology,
        Tokyo University of Science, 2641, Noda, Chiba, 278-8510, Japan,
       email: omori@ma.noda.tus.ac.jp}
        \\ Tokyo University of Science
\and  Yoshiaki Maeda\thanks{Department of Mathematics,
                Faculty of Science and Technology,
                Keio University, 3-14-1, Hiyoshi, Yokohama,223-8522, Japan,
                email: maeda@math.keio.ac.jp}
          \\Keio University
\and  Naoya Miyazaki\thanks{ Department of Mathematics, Faculty of
Economics, Keio University,  4-1-1, Hiyoshi, Yokohama, 223-8521, Japan,
        email: miyazaki@hc.cc.keio.ac.jp}
        \\Keio University
\and  Akira Yoshioka \thanks{ Department of Mathematics,
          Faculty of Science, Tokyo University of Science,
         1-3, Kagurazaka, Tokyo, 102-8601, Japan,
         email: yoshioka@rs.kagu.tus.ac.jp}
           \\Tokyo University of Science
     }

\maketitle



\setcounter{equation}{0}

\tableofcontents

\bigskip

This paper presents a preliminary version of the deformation theory of 
expressions of elements of algebras. 
The notion of $*$-functions is given.  
Several important problems appear in  simplified forms, 
and these give an intuitive bird's-eye of the whole theory, since  
these problems are main target of this series. 
These will give a toy model
of the philosophical  ``the theory of observations''.

This series will continue to twelve chapters at least. 

\section{Definition of $*$-functions and intertwiners}
Let ${\mathbb C}[w]$ be the space of polynomials in 
one variable $w$. For a complex parameter $\tau$,
we define a new product $*_{\tau}$ on this space 
\begin{equation}
 \label{eq:prodgen}
f{*}_{\tau}g=\sum_{k\geq 0}
\frac{\tau^k}{2^kk!}\partial^k_{w}f\partial^k_{w}g  
\quad(\,\,=fe^{\frac{\tau}{2}
\overleftarrow{\partial_{{w}}}\overrightarrow{\partial_{{w}}}}g). 
\end{equation} 
We easily see that the  $*_{\tau}$ makes 
${\mathbb C}[w]$ a commutative associative algebra, 
which we denote by 
$({\mathbb C}[w], *_{\tau})$. If $\tau{=}0$, then 
$({\mathbb C}[w], *_{0})$ is the usual polynomial algebra, and 
$\tau\in {\mathbb C}$ is called a deformation parameter.
What is deformed is not the algebraic structure, 
but the expression of elements. 

We see easily that
$w{*_{\tau}}w=w^2{+}\frac{\tau}{2}$, \,\,
$w{*_{\tau}}w{*_{\tau}}w=w^3{+}\frac{3\tau}{2}w$, etc. 

\subsection{Intertwiners and  infinitesimal intertwiners}

It is not hard to verify that the mapping  
\begin{equation}
 \label{eq:intwnner}
e^{\frac{\tau}{4}\partial^2_{w}}: ({\mathbb C}[w],{*}_{0})\to
({\mathbb C}[w],{*}_{\tau})   
\end{equation}
gives an algebra isomorphism. That is, 
$e^{\frac{\tau}{4}\partial^2_{w}}$ has the inverse mapping   
$e^{-\frac{\tau}{4}\partial^2_{w}}$ and 
the following holds
\begin{equation}
  \label{eq:isoiso}
 e^{\frac{\tau}{4}\partial^2_{w}}(f{*}_0g){=}
(e^{\frac{\tau}{4}\partial^2_{w}}f){*}_{\tau}
(e^{\frac{\tau}{4}\partial^2_{w}}g). 
\end{equation}
This isomorphism 
$I_0^{\tau}{=}e^{\frac{\tau}{4}\partial^2_{w}}$ 
is called the {\bf intertwiner}.  
We define also 
$I_{\tau}^{\tau'}=I_0^{\tau'}(I_0^{\tau})^{-1}$ 
as the intertwiner from $*_\tau$ onto ${*_{\tau'}}$, and its differential 
$$
dI_{\tau}= I_{\tau}^{\tau{+}d\tau}=
\frac{d}{d\tau'}I_{\tau}^{\tau'}\Big|_{\tau'=\tau}=\frac{1}{4}\partial^2_{w}
$$
is called the {\bf infinitesimal intertwiner}.

\medskip
Intertwiners are powerful tool to compute ${*_{\tau}}$-products. 
We compute $w_{*\tau}^n$ by $I_0^{\tau}w^n$:   
\begin{equation}\label{Pn00}
w_{*\tau}^n{=}
P_n(w,\tau){=}
\sum_{k\leq [n/2]}\frac{n!}{4^kk!(n{-}2k)!}\tau^kw^{n{-}2k}.
\end{equation}

For the later use, we note that intertwiners leaves the {\it parity}
invariant. That is,
\begin{prop}\label{evenoddprev}
$w_{*\tau}^{2n}$ is written by a polynomial $p(w^2)$ of $w^2$, and 
$w_{*\tau}^{2n{+}1}$ is a polynomial written in the form $w\,p(w^2)$.
\end{prop}

\medskip
Let $H\!ol({\mathbb C})$ be the space of all entire functions on 
${\mathbb C}$ with the topology of uniform convergence 
on compact domains. 
$H\!ol({\mathbb C})$ is known to be a Fr{\'e}chet space 
defined by a countable family of seminorms.
It is easy to see that the product $*_{\tau}$ extends naturally for  
$f, g\in H\!ol({\mathbb C})$ if $f$ or $g$ is a polynomial. 
We easily see the following: 
\begin{thm}\label{fund00pol}
For every polynomial $p(w)$, the multiplication operator  
$p(w){*_\tau}$  is a continuous linear mapping 
of $H\!ol({\mathbb C})$ into itself. 

By polynomial approximations, the associativity 
$f{*_\tau}(g{*_\tau}h)=(f{*_\tau}g){*_\tau}h$ holds 
if two of $f,g,h$ are polynomials.   

By giving the inductive limit topology, ${\mathbb C}[w]$ 
is a complete topological algebra, and 
$H\!ol({\mathbb C})$ is a topological 
${\mathbb C}[w]$ bi-module.
\end{thm}

By Theorem\,\ref{fund00pol}, every polynomial $p(w)$ may be 
viewed as a linear operator of $H\!ol({\mathbb C})$ into itself. 
Thus, for instance, 
one may consider the differential equation 
$$
\frac{d}{dt}f_t(w)=w_*^{\ell}{*}f_t(w), \quad f_0(w)=1.
$$
In ordinary calculus, this is treated by using 
slightly transcendental elements  
$\sum_k \frac{t^{k}}{k!}w_*^{n\ell}$ via the notion 
of convergence, which is familiar in the ordinary text book 
of calculus, and there is no problem for the case $\ell=1$.

\subsection{$*$-exponential functions and 
$\tau$-expressions}\label{111}

Although the ordinary exponential function $e^{aw}$ is not 
a polynomial, the intertwiner extends to give a family of elements   
\begin{equation}
  \label{eq:expdef}
 I_0^{\tau}(e^{2aw}){=}e^{2aw{+}a^2\tau}{=}
e^{a^2\tau}e^{2aw}, \quad \tau\in{\mathbb C}
\end{equation}
and Taylor expansion formula gives 
\begin{equation}
\label{eq:prodexp}
e^{2aw}{*_{\tau}}e^{2bw}{=}
e^{2(a{+}b)w{+}2ab\tau},\quad
e^{2aw}{*_{\tau}}f(w)=e^{2aw}f(w{+}a\tau)
\end{equation}
for every $f{\in}H\!ol({\mathbb C})$. 
Using this, we see the associativity 
\begin{equation}\label{assocwrtexp}
e^{2aw}{*_\tau}(e^{2bw}{*_\tau}f(w))
{=}(e^{2a\tau}{*_\tau}e^{2bw}){*_\tau}f(w)
\end{equation}
for every $f{\in}H\!ol(\mathbb C)$.   
Computation via intertwiners gives     
$$
I_{\tau}^{\tau'}(e^{s{w}}){=}
e^{\frac{1}{4}(\tau'{-}\tau)s^2}e^{s{w}}, \quad    
I^{\tau'}_{\tau}(e^{\frac{1}{4}s^2\tau}e^{s{w}})
{=}e^{\frac{1}{4}s^2\tau'}e^{s{w}}.
$$
Noting \eqref{eq:expdef}, we write the family 
$\{e^{\frac{1}{4}s^2\tau}e^{s{w}}; \tau\in{\mathbb C}\}$ by 
$e_*^{s{w}}$ and call this the $*$-{\bf exponential function}. 

\subsubsection{Remarks for the notations}

In what follows we use notations such as 
$$
{:}e_*^{sw}{:}_{\tau}=e^{\frac{1}{4}s^2\tau}e^{s{w}}
$$ 
and call the r.h.s. the $\tau$-expression of $e_*^{s{w}}$. 
Similarly we denote ${:}w_*^n{:}_{\tau}=P_n(w,\tau)$ and call the
r.h.s. the $\tau$-expression of $w_*^n$.
  
In general, for a polynomial or an exponential function $f(w)$, 
there is a family of functions  
$$
\{f_{{\tau}}(w); \tau{\in}{\mathbb C}\}, \quad f_{\tau}{=}I_0^{\tau}(f).
$$ 
We denote this family 
by $f_*(w)$ and we view $f_*(w)$ as an {\it element} of 
the abstract algebra. We refer to such objects as a 
$*$-{\bf function}s. 
Furthermore, we denote these as 
\begin{equation}
  \label{eq:tempexp}
{:}f_*(w){:}_\tau{=}f_\tau(w) 
\end{equation}
by using the notation ${:}\btt{:}_{\tau}$. 
${:}f_*{:}_{\tau}$ is 
viewed as the $\tau$-{\bf expression} of $f_*$. 
For instance, we  write 
$$
{:}a{w}_*{+}b{:}_{\tau}=aw{+}b, \quad 
{:}2{w}_*^2{:}_{\tau}=2{w}^2{+}\tau,\quad 
{:}2{w}_*^3{:}_{\tau}=2{w}^3{+}3\tau {w},\quad 
{:}{w}_*^n{:}_{\tau}=P_n(w,\tau)\,\, ({\text{cf.}}\eqref{Pn00}). 
$$

\bigskip
 For $*$-functions 
$f_*(w), g_*(w)$, the notations $f_*(w){*}g_*(w)$ and 
$f_*(w)+g_*(w)$ without expression parameter signs 
mean that these are computed under the same expression 
parameter. The usual calculations will be done in this manner. 
Calculations such as 
$$
{:}f_*(w){:}_{\tau}+{:}g_*(w){:}_{\tau'},\quad 
{:}f_*(w){:}_{\tau}{:}g_*(w){:}_{\tau'}
$$
are forbidden, though these are calculations in ordinary functions.

\bigskip
It is sometimes very convenient to regard the infinitesimal intertwiner 
as a linear connection on the trivial bundle 
$\coprod_{\tau\in{\mathbb C}}H{\!o}l(\mathbb C)$. 
But, the parallel transformation do no necessarily defined. 
If $I_0^{\tau}f$ is defined on some domain $D$,  
this is viewed as a parallel section defined on $D$. 
In the later section, we treat multi-valued parallel sections 
if $D$ is not simply connected.

As the intertwiner $I_0^{\tau}$ is an isomorphism on the space of
polynomials, the following is easy to see: 
\begin{cor}\label{ufacthm}
The unique factorization holds for 
$*$-polynomials $p_*(w)$, i.e. $p_*(w)$ is written uniquely in the
form 
$$
p_*(w)=a_0(w{+}\lambda_1)^{\ell_1}{*}(w{+}\lambda_2)^{\ell_2}{*}\cdots{*}(w{+}\lambda_k)^{\ell_k}.
$$
\end{cor}

By the product formula \eqref{eq:prodgen} we have 
the exponential law
\begin{equation}
  \label{eq:explaw1}
{:}e_*^{s{w}}{:}_{\tau}{*_{\tau}}{:}e_*^{t{w}}{:}_{\tau}=
{:}e_*^{(s{+}t){w}}{:}_{\tau},
\quad \forall \tau\in {\mathbb C}.  
\end{equation}
Furthermore 
${:}e_*^{t{w}}{:}_{\tau}$ is the solution of the differential equation  
$\frac{d}{dt}g(t)={w}{*_{\tau}}g(t)$ with the initial condition 
$g(0){=}1$ for every $\tau$.
By the uniqueness of solutions, the exponential law with the ordinary exponential
function $e^{s}$
$$
e_*^{tw}e^{s}=e_*^{tw{+}s}
$$
holds. As $I_0^{\tau}(e^{tw})={:}e_*^{tw}{:}_{\tau}$ and 
$I_0^{\tau}(w^n){=}{:}w_*^n{:}_{\tau}$, 
we see that 
$$
{:}e_*^{tw}{:}_\tau{=}\sum_n \frac{t^n}{n!}{:}w_*^n{:}_\tau.
$$

Thus, it is natural to view 
$e_*^{s{w}}$ as a single element, rather than a 
parallel section defined on the whole plane ${\mathbb C}$, or a  
family of elements which are mutually intertwined.

The exponential law is written as   
$e_*^{s{w}}{*}e_*^{t{w}}=e_*^{(s{+}t){w}}$,  
and the equation itself is written as 
$\frac{d}{dt}g_*(t)=w{*}g_*(t)$, $g_*(0){=}1$ 
without the subscript $\tau$. 

\medskip
We define naturally 
$$
\sin_*(w{+}s)=\frac{1}{2i}(e_*^{i(w{+}s)}{-}e_*^{-i(w{+}s)}),\quad 
\cos_*(w{+}s)=\frac{1}{2}(e_*^{i(w{+}s)}{+}e_*^{-i(w{+}s)}).
$$

\medskip

\begin{lem}
\label{ketugoprodexp22}  
For every $f\in H\!ol(\mathbb C)$,  
the formula \eqref{eq:prodexp} gives 
\begin{equation} 
 \label{eq:prodexp22}
{:}e_*^{2s{w}}{:}_\tau{*_\tau}f({w})
=e^{2s{w}{+}s^2\tau}f({w}{+}s\tau)   
\end{equation} 
and it holds the following associativity in every 
$\tau$-expression 
$$
e_*^{rw}{*}(e_*^{sw}{*}f){=}e_*^{(r{+}s)w}{*}f
{=}e_*^{rw}{*}(f{*}e_*^{sw}).
$$
\end{lem}

As ${:}e_*^{2nw}{:}_{\tau}=e^{n^2\tau}e^{2nw}$, 
if ${\rm{Re}}\tau{<}0$, then 
we easily see that $\lim_{n\to\infty}{:}e_*^{2nw}{:}_{\tau}=0$ 
uniformly on every compact domain. 
However, this does not necessarily imply that 
$\lim_{n\to\infty}{:}e_*^{2nw}{*}f{:}_{\tau}=0$, for 
\eqref{eq:prodexp22} gives
$$
\lim_{n\to\infty}{:}e_*^{2nw}{:}_{\tau}*_{\tau}f(w)=
\lim_{n\to\infty} e^{n^2\tau{+}2nw}f(w+n\tau).
$$

\noindent
{\bf Note}\,\,So far, the generator $w$ is fixed.  
Note here that a change of generators 
$w\to w'{=}aw$ causes often a confusion, since
\begin{equation}\label{genchnge}
{:}e_*^{(ta)w}{:}_{\tau}=e^{\frac{1}{4}(ta)^2\tau}e^{(ta)w} 
\not= 
{:}e_*^{t(w')}{:}_{\tau}{=}
e^{\frac{1}{4}t^2\tau}e^{tw'}.
\end{equation}
The effect of changing generator systems will be discussed 
in the forthcoming paper. Here we note that 
a change of an expression parameter is recovered by  
the ``square'' of a change of generators. 
Thus a change of expression parameter may be viewed as a 
``square root'' of a coordinate transformation. 

\bigskip
\noindent
{\bf Remark}\,\, By definition of $*$-exponential function, 
we have 
$e_*^{0w}=1$ for every $\tau$-expression. 
This notation is a little confusing, for it does not imply 
${:}e_*^{aw}{:}_{\tau}\big|_{w=0}=1$. 
Instead, it is computed as 
$$
{:}e_*^{aw}{:}_{\tau}\big|_{w=0}=
e^{\frac{1}{4}a^2\tau}e^{a0}=e^{\frac{1}{4}a^2\tau}.
$$
Hence the evaluation $\big|_{w=0}$ is not a homomorphism: 
$$
{:}e_*^{aw}{*}{e_*^{bw}:}_{\tau}\big|_{w=0}=
e^{\frac{1}{4}(a{+}b)^2\tau}\not=e^{\frac{1}{4}a^2\tau}e^{\frac{1}{4}b^2\tau}.
$$

\begin{center}
\begin{minipage}{.85\linewidth}
Star-functions of one variable $w$ may be viewed as 
simplified version of star-functions of a linear function 
$\langle{\pmb a},{\pmb u}\rangle$ of 
several variables for a fixed $\pmb a$. 
\end{minipage}
\end{center}

\bigskip
So far, we treated the differential equation   
$$
\frac{d}{dt}f_t(w)=w_*^{\ell}{*}f_t(w), \quad f_0(w)=1
$$
in the case $\ell=1$. The solution is given by 
$\sum_k \frac{t^{k}}{k!}w_*^{k}$.
 The case $\ell=2$ will be treated  in the later section. 
It will be seen that $e_*^{tw_*^2}$ behaves 
{\it 2-valued complex one parameter group}  
with a branching singular point at $\tau^{-1}$. 

\medskip
In general, by noting that ${:}w_{*}^{\ell n}{:}_{\tau}=w_{*\tau}^{\ell n}$,
 \eqref{Pn00} gives that 
$$
\sum_n\frac{t^n}{n!}{:}w_{*}^{\ell n}{:}_{\tau}=
\sum_{m,k\geq 0}\frac{(m{+}2k)!}{n!}
\frac{1}{k!m!}t^{n}(\frac{\tau}{4})^kw^m,\quad
\ell n=m{+}2k.
$$
Set the r.h.s. by 
$$
\sum_mA_m(t,\tau)\frac{1}{m!}w^m,\quad 
A_m(t,\tau)=\sum_{k}\frac{(m{+}2k)!}{([\frac{1}{\ell}(m{+}2k)])!}
t^{[\frac{1}{\ell}(m{+}2k)]}\frac{1}{k!}(\frac{\tau}{4})^k
$$
where summation runs through all $k$ such that $\ell n=m{+}2k$.
$A_m(t,\tau)$ is a formal power series  
$A_m(t,\tau)=\sum_kA_{m,k}(t^{\frac{1}{\ell}})\frac{1}{k!}(\frac{\tau}{4})^k$
of $\tau$.  
By this we have the following:
\begin{prop}\label{convzero}
If $\ell\geq 3$, then for every $t\not=0$, the radius of convergence of the power
series  
$\sum_{k=0}^{\infty}A_{m,k}(t^{\frac{1}{\ell}})\frac{1}{k!}(\frac{\tau}{4})^k$
 is $0$. 
That is, 
$e_*^{tw_*^{\ell}}$ cannot be defined as a power series for 
$\ell\geq 3$.
\end{prop}

\noindent
{\bf Proof}\,\,
Note that the lowest term of 
$P_{\ell n}(w,\tau)$ is 
$\frac{(\ell n)!}{4^{\ell n/2}(\ell n/2)!}\tau^{\ell n/2}$ or 
$\frac{(\ell n)!}{4^{[\ell n/2]}([\ell n/2])!}
\tau^{[\ell n/2]}w$. Denote this by $a_n$.
If $\ell\geq 3$, then 
$$
\lim_{n\to\infty}\frac{(n+1)|a_n|}{|a_{n+1}|}=0.
$$
This means the radius of convergence of $A_{0}(t,\tau)$ is $0$. Others
are similarly. 
${}$\hfill $\Box$

In the later section,  
$e_*^{tw_*^{\ell}}$, $\ell\geq 3$, is defined as a certain object 
which has some similarity to a  
{\it multi-valued complex one parameter group}  
with a branching singular point at $t=0$.

\subsection{Applications to generating functions} 

As in the ordinary calculus, several generating functions are obtained
by exponential functions. In this section, we apply these to
$*$-exponential functions to obtain slight generalizations.

\subsubsection{The generating function of Hermite polynomials} 

The generating function of Hermite polynomials is given as follows:
$$
e^{\sqrt{2}tx{-}\frac{1}{2}t^2}
=\sum_{n=0}^\infty H_n(x)\frac{t^n}{n!} .
$$
This is the Taylor expansion formula of 
${:}e_*^{\sqrt{2}tw}{:}_{-1}$. 

Noting that $e_*^{taw}=\sum\frac{t^n}{n!}(aw)_*^n$, and setting 
$$
e_*^{\sqrt{2}tw}{=}\sum_{n\geq 0}(\sqrt{2}w)_*^n\frac{t^n}{n!},
$$
we see $H_n(w)={:}(\sqrt{2}w)_*^n{:}_{-1}$. Note that $H_n(w)$ is a polynomial of degree $n$.
For every $\tau\in \mathbb C$ we define $*$-Hermite polynomials $H_n(w,*)$ by 
\begin{equation}\label{hermite}
e_*^{\sqrt{2}tw}{=}\sum_{n\geq 0}^\infty H_n(w,*)\frac{t^n}{n!},
\qquad (H_n(w,\tau){=}{:}H(w,*){:}_{\tau},\quad H_n(w,-1){=}H_n(w)) .
\end{equation} 
Since 
$\frac{d}{dt}e_*^{\sqrt{2}tw}{=}\sqrt{2}w{*}e_*^{\sqrt{2}tw}$, 
we have 
\begin{equation}\label{hermite00}
\frac{\tau}{\sqrt{2}}H'_n(w,\tau){+}\sqrt{2}wH_n(w,\tau)=H_{n{+}1}(w,\tau)
\end{equation}
where $H'_n(w,\tau){=}\frac{\partial}{\partial w}H(w,\tau){=}
{:}\frac{\partial}{\partial w}H(w,*){:}_{\tau}$. 
The exponential law yields 
$$
\sum_{k+\ell=n}\frac{n!}{k!{\ell}!}H_k(w,*){*}H_{\ell}(w,*)=
H_n(w,*).
$$
On the other hand, taking 
$\frac{\partial}{\partial w}$ of both sides of 
\eqref{hermite} gives 
$$
\sqrt{2}nH_{n{-}1}(w,*)=H'_n(w,*).
$$
Differentiating \eqref{hermite00} again 
and using the above equality gives  
$$
{\tau}H''_n(w,\tau){+}2wH'_n(w,\tau){-}2nH_{n}(w,\tau)=0.
$$

By  
$\sqrt{2}tw+\frac{\tau}{2}t^2=
\frac{\tau}{2}(t{+}\frac{\sqrt{2}}{\tau}w)^2-\frac{1}{\tau}w^2$,
the Hermite polynomial $H_n(w,*)$ is obtained by the following formula:
$$
\begin{aligned}
H_n(w,\tau)&{=}
\frac{d^n}{dt^n}{:}e_*^{\sqrt{2}tw}{:}_{\tau}\Big|_{t=0}=
\frac{d^n}{dt^n}
e^{\frac{\tau}{2}(t{+}\frac{\sqrt{2}}{\tau}w)^2-\frac{1}{\tau}w^2}\Big|_{t=0}\\
&=
\frac{d^n}{dt^n}
e^{\frac{\tau}{2}(t{+}\frac{\sqrt{2}}{\tau}w)^2}\Big|_{t=0}e^{-\frac{1}{\tau}w^2}
=e^{-\frac{1}{\tau}w^2}(\frac{\tau}{\sqrt{2}})^n\frac{d^n}{dw^n}e^{\frac{1}{\tau}{w^2}}
\end{aligned}
$$

To obtain the orthogonality of the family $\{H_n(w,\tau)\}_n$, we suppose
${\rm{Re}}\tau<0$ and we restrict $w$ to the real line.
Then, the orthogonality is shown as follows:
$$
\int_{\mathbb R}e^{\frac{1}{\tau}w^2}H_n(w,\tau)H_m(w,\tau)dw=
\int_{\mathbb R}(\frac{\tau}{\sqrt{2}})^n\frac{d^n}{dw^n}
e^{\frac{1}{\tau}{w^2}} H_m(w,\tau)dw.
$$
If $n\not= m$, one may suppose $n>m$ without loss of generality. 
Hence this vanishes by integration by parts $n$ times.
For the case $n=m$, since   
$$
{:}e_*^{\sqrt{2}tw}{:}_{\tau}{=}e^{\frac{\tau}{2}t^2{+}\sqrt{2}tw}
=\sum_{n=0}^\infty H_n(w,\tau)\frac{t^n}{n!} 
$$ 
 by \eqref{hermite}, we see 
$$
\frac{1}{n!}H_n(w,\tau)=\sum_{p=0}^{[n/2]}\frac{\sqrt{2}^n\tau^p}{p!(n-2p)!4^p}w^{n-2p}
$$
and hence 
$$
\frac{d^n}{dw^n}H_n(w,\tau)=\sqrt{2}^nn!. 
$$
It follows 
$$
\int_{\mathbb R}e^{\frac{1}{\tau}w^2}H_n(w,\tau)H_n(w,\tau)dw=
n!(-\tau)^n\int_{\mathbb R}e^{\frac{1}{\tau}w^2}dw=
n!(-\tau)^n\sqrt{-\tau}\sqrt{\pi}. 
$$

\subsubsection{The generating function of Legendre polynomials}

The generating function of Legendre polynomials $P_n(z)$ is known 
to be 
$$
\frac{1}{\sqrt{1{-}2tz{+}t^2}}=\sum_{n=0}^{\infty}P_n(z)t^n, \quad
\text{for small }\,\,|t|.
$$
It is known that 
$P_n(z)=\frac{1}{2^nn!}\frac{d^n}{dz^n}(z^2{-}1)^{n}$.
Hence 
\begin{equation}\label{LegTaylor}
\frac{1}{\sqrt{1{-}2t(z{+}a){+}t^2}}=\sum_n\frac{1}{2^nn!}\frac{d^n}{da^n}((z{+}a)^2{-}1)^{n}t^n
\end{equation}
is viewed as the Taylor expansion of the l.h.s. of \eqref{LegTaylor}. 
By the formula of Laplace transform, rewrite the l.h.s. of \eqref{LegTaylor}, and we see   
$$
\frac{1}{\sqrt{1{-}2t(z{+}a){+}t^2}}
=
\frac{1}{\sqrt{\pi}}
\int_0^{\infty}\frac{1}{\sqrt{s}}e^{-s(1{-}2t(z{+}a){+}t^2)}ds =
\sum_{n=0}^{\infty}P_n(z{+}a)t^n.
$$ 
This implies also that  
\begin{equation}\label{laplace}
\frac{d^n}{dt^n}\Big|_{t=0}\frac{1}{\sqrt{\pi}}
\int_0^{\infty}\frac{1}{\sqrt{s}}e^{-s(1{-}2t(z{+}a){+}t^2)}ds
=\frac{1}{2^n}\frac{d^n}{da^n}((z{+}a)^2{-}1)^{n}.
\end{equation}

Replacing the exponential function in the integrand by the
$*$-exponential function, we define $*$-Legrendre polynomial by 
$$
\frac{1}{\sqrt{\pi}}
\int_0^{\infty}\frac{1}{\sqrt{s}}e_*^{-s(1{-}2t(w{+}a){+}t^2)}ds =
\sum_{n=0}^{\infty}P_n(w{+}a,*)t^n.
$$
As ${:}e_*^{-s(1{-}2t(w{+}a){+}t^2)}{:}_{\tau}=e^{\tau s^2t^2}e^{-s(1{-}2t(w{+}a){+}t^2)}$,
we assume that ${\rm{Re}}\,\tau<0$ so that the integral converges.
$$
\frac{1}{\sqrt{\pi}}
\int_0^{\infty}\frac{1}{\sqrt{s}}e^{\tau s^2t^2}e^{-s(1{-}2t(w{+}a){+}t^2)}ds =
\sum_{n=0}^{\infty}P_n(w{+}a,\tau)t^n,\quad P_n(w{+}a,\tau)={:}P_n(w{+}a,*){:}_{\tau}.
$$
As the variable $z$ is used formally in \eqref{laplace}, the same
formula as in \eqref{laplace} holds in $*$-exponential functions: 
That is, 
$$
\frac{d^n}{dt^n}\Big|_{t=0}\frac{1}{\sqrt{\pi}}
\int_0^{\infty}\frac{1}{\sqrt{s}}e_*^{-s(1{-}2t(w{+}a){+}t^2)}ds
=\frac{1}{2^n}\frac{d^n}{da^n}((w{+}a)_*^2{-}1)_*^{n}.
$$
By this trick we see that 
\begin{equation}\label{proofortho}
\frac{1}{\sqrt{\pi}}
\int_0^{\infty}\frac{1}{\sqrt{s}}e_*^{-s(1{-}2t(w{+}a){+}t^2)}ds
=
\sum_{n=0}^{\infty}P_n(w{+}a,*)t^n
=\sum_n\frac{1}{2^nn!}\frac{d^n}{da^n}((w{+}a)_*^2{-}1)_*^{n}t^n. 
\end{equation}
Hence we see that 
$P_n(w{+}a,*)$ is a polynomial of degree $n$ and 
$\frac{d^n}{da^n}P_n(w{+}a,*)=(2n{-}1)!!$.
This equality is used to obtain the orthogonality of the family
$\{P_n(w{+}a),\tau\}_n$. We restrict $w{+}a$ to the real line.
Integration by parts gives  
$$
\int_{-1}^{1}P_m(w,\tau)P_n(w,\tau)dw=\frac{2}{2n{+}1}\delta_{m,n}.
$$

\subsubsection{The generating function of Bessel functions} 

The generating function of Bessel functions $J_n(z)$ is known to be 
$$
e^{iz\sin s}= \sum_{n=-\infty}^{\infty}J_n(z)e^{ins}.
$$
Keeping this in mind, we define $*$-Bessel functions by 
$$
e_*^{\frac{1}{2}(e^{is}{-}e^{-is})aw}= \sum_{n=-\infty}^{\infty}J_n(aw,*)e^{ins}, \quad 
{:}J_n(aw,*){:}_{\tau}=J_n(aw,\tau),\quad a\in{\mathbb C}.
$$
Replacing $s$ by $s{+}\frac{\pi}{2}$ gives 
$
e_*^{\frac{i}{2}(e^{is}{+}e^{-is})aw}=\sum_{n=-\infty}^{\infty}i^nJ_n(aw,*)e^{ins}
$
and basic symmetric properties hold: 

First we see 
$J_n(aw,*)=(-1)^nJ_{-n}(aw,*)$. Replacing $w$ by $-w$ in the first
equality gives $J_n(-aw,*)=J_{-n}(aw,*)$.    
Since 
$$
{:}e_*^{\frac{1}{2}(e^{is}{-}e^{-is})aw}{:}_{\tau}=
e^{\frac{-a^2}{16}\tau(e^{is}{-}e^{-is})^2}e^{\frac{1}{2}(e^{is}{-}e^{-is})aw}
=e^{\frac{-a^2}{8}\tau}
e^{\frac{a^2}{16}\tau(e^{2is}{+}e^{-2is})}e^{\frac{1}{2}(e^{is}{-}e^{-is})aw},
$$
$J_n(aw,\tau)$ and $J_n(aw)$ are related by 
$$
\sum_{n=-\infty}^{\infty}J_n(aw,\tau)e^{ins}=
e^{-\frac{a^2}{8}\tau}
e^{\frac{a^2}{16}\tau(e^{2is}{+}e^{-2is})}
\sum_{n=-\infty}^{\infty}J_n(aw)e^{ins}.
$$
Setting $s=0$, we see in particular 
$$
1=\sum_{n=-\infty}^{\infty}J_n(aw,\tau)=\sum_{n=-\infty}^{\infty}J_n(aw).
$$

The exponential law of l.h.s. of the defining equality gives that 
$$
e_*^{\frac{1}{2}(e^{is}{-}e^{-is})aw}{*}e_*^{\frac{1}{2}(e^{is}{-}e^{-is})bw}
=e_*^{\frac{1}{2}(e^{is}{-}e^{-is})(a{+}b)w}=
\sum_nJ_n(aw{+}bw,*)e^{ins}.
$$
Hence, 
$$
J_n(aw{+}bw,*)=\sum_{m=-\infty}^{\infty}J_m(aw,*){*}J_{n{-}m}(bw). 
$$

If $a^2{+}b^2=1$, then  
$$
e_*^{\frac{1}{2}(e^{is}{-}e^{-is})aw}{*}e_*^{\frac{i}{2}(e^{is}{+}e^{-is})bw}
=e_*^{\frac{1}{2}((a{+}ib)e^{is}{-}(a{-}ib)e^{-is})w}=
\sum_nJ_n(w,*)(a{+}ib)^ne^{nis}.
$$ 
$$
\sum_{k=-\infty}^{\infty}J_k(aw,*)e^{iks}{*}\sum_{\ell=
-\infty}^{\infty}{i^\ell}J_\ell(bw,*)e^{i{\ell}s}
{=}\sum_nJ_n(w,*)(a{+}ib)^ne^{nis}.
$$

\bigskip
The generating functions of Bernoulli numbers, Euler numbers  
and Laguerre polynomials  will be treated later
sections, for there are some other problems for the 
treatment. 
  
\subsection{Jacobi's theta functions and Imaginary transformations} 

For arbitrary $a{\in}{\mathbb C}$, consider the 
$*$-exponential function 
$e_*^{t(w+a)}$. Since 
$$
{:}e_*^{t(w+a)}{:}_{\tau}=e^{\frac{\tau}{4}t^2}e^{t(w+a)},
$$
by assuming $\rm{Re}\,\tau <0$, this is rapidly decreasing on
${\mathbb R}$. Hence, we see that both 
$$
\int_{-\infty}^{\infty}{:}e_*^{t(w+a)}{:}_{\tau}dt,\quad 
\sum_{n=-\infty}^{\infty}{:}e_*^{n(w+a)}{:}_{\tau}
$$
converge absolutely on every compact domain in $w$ 
to give entire functions of $w$. 

In this section, we treat first a special case
$\theta(w,*)=\sum_ne_*^{2inw}$
under the condition  
$\rm{Re}\,\tau >0$.
If we set $q{=}e^{-\tau}$, the $\tau$-expression 
$\theta(w,\tau){=}{:}{\theta}(w,*){:}_{\tau}$ is given by  
$$
\theta(w,\tau){=}\sum_{n{\in}{\mathbb Z}}q^{n^2}e^{2niw}. 
$$
This is Jacobi's elliptic $\theta$-function $\theta_3(w,\tau)$.

Furthermore, Jacobi's 
elliptic theta functions $\theta_i, i{=}1,2,3,4$ are all 
$\tau$-expressions of bilateral geometric series as 
follows (cf. \cite{AAR}).
\begin{equation}
 \label{eq:jacobi}
\begin{aligned}
{\theta}_1(w,*)=&
\frac{1}{i}\sum_{n=-\infty}^{\infty}(-1)^ne_*^{(2n{+}1)iw},\qquad 
{\theta}_2(w,*)=  \sum_{n=-\infty}^{\infty}e_*^{(2n{+}1)iw},\\
{\theta}_3(w,*)=&
 \sum_{n=-\infty}^{\infty}e_*^{2niw},\qquad\qquad\qquad\quad
{\theta}_4(w,*)=\sum_{n=-\infty}^{\infty}(-1)^ne_*^{2niw}
\end{aligned}  
\end{equation}

\smallskip 
This fact was mentioned first 
in \cite{om6}, but no further investigation of this fact 
has been done. 

By the exponential law $e_*^{aw{+}s}=e_*^{aw}e^{s}$ for $s{\in}
{\mathbb C}$, which is proved easily, we see that  
${\theta}_i(w,*)$ is $2\pi$-periodic for every $i$. (Precisely,
$\theta_1(w,*)$, $\theta_2(w,*)$ are alternating $\pi$-periodic, and 
$\theta_3(w,*)$, $\theta_4(w,*)$ are $\pi$-periodic.)  
Furthermore the exponential law 
$e_*^{aw{+}bw}=e_*^{aw}{*}e_*^{bw}$ of \eqref{eq:explaw1}
gives the trivial identities 
$$
e_*^{2iw}{*}{\theta}_i(w,*)={\theta}_i(w,*),\,\, (i{=}2,3),
\quad
e_*^{2iw}{*}{\theta}_i(w,*)=-{\theta}_i(w,*),\,\, (i{=}1,4).
$$

For every $\tau$ such that ${\rm{Re}}\,\tau >0$, 
$\tau$-expressions of these are given by using 
${:}e_*^{2iw}{:}_{\tau}=e^{-\tau}e^{2iw}$ and 
\eqref{eq:prodexp22} as follows:
\begin{equation}
\begin{aligned}
&e^{2iw-\tau}\theta_i(w{+}i\tau, \tau){=}\theta_i(w,\tau),
 \,\,(i=2,3),\\ 
&e^{2iw-\tau}\theta_i(w{+}i\tau, \tau){=}{-}\theta_i(w,\tau), \,\,(i=1,4).
\end{aligned}
\end{equation}
$\theta_i(w;*)$ is a parallel section defined on the open 
right half-plane of the expression parameters, but 
the expression parameter $\tau$ turns out to give  
quasi-periodicity with the exponential factor $e^{2iw-\tau}$. 

Noting that $(e_*^{2iw}{-}1){*}\theta_3(w, *){=}0$ in the 
computation of $*_{\tau}$-product, we have  
\begin{prop}\label{easy00}
If $f{\in}H\!ol(\mathbb C)$ satisfies $f(w{+}\pi){=}f(w)$ and
${:}(e_*^{2iw}{-}1){:}_\tau{*_\tau}f{=}0$, then 
$$
f{=}c{:}\theta_3(w, *){:}_\tau,\,\, c{\in}{\mathbb C}.
$$
\end{prop} 

\noindent
{\bf Proof}\, By periodicity, the Fourier expansion theorem gives 
$f(w){=}\sum a_ne^{2in w}$, but by the formula of $*$-exponential functions, this is 
rewritten as $f(w)={:}\sum c_ne_*^{2in w}{:}_\tau$. 
This gives the result, for the second identity gives 
that $c_{n+1}=c_n$. \hfill $\square$

\subsubsection{Two different inverses of an element}\label{sect-invss}

The trivial identities 
of theta functions give many inverses. The convergence of bilateral 
geometric series gives a little strange feature. Note that if 
${\rm{Re}}\,\tau>0$, then $\tau$-expressions of  
$\sum_{n=0}^{\infty}e_*^{2niw}$ and    
$-\sum_{n=-\infty}^{-1}e_*^{2niw}$ both converge in $H\!ol(\mathbb C)$ 
to give inverses of the element ${:}(1{-}e_*^{2iw}){:}_{\tau}$, and 
$\theta_3(w,\tau)$ is the difference of these inverses. 

We denote these inverses by using short notations:
$$
(1{-}e_*^{2iw})^{-1}_{*+}{=}\sum_{n=0}^{\infty}e_*^{2niw},\quad
(1{-}e_*^{2iw})^{-1}_{*-}{=}-\sum_{n=1}^{\infty}e_*^{-2niw},\quad 
(1{-}e_*^{-2iw})^{-1}_{*+}{=}\sum_{n=1}^{\infty}e_*^{-2niw}
$$
Apparently, this breaks associativity:
$$
\Big((1{-}e_*^{2iw})^{-1}_{*+}{*_\tau}(1{-}e_*^{2iw})\Big){*_\tau}(1{-}e_*^{2iw})^{-1}_{*-}
\not=
(1{-}e_*^{2iw})^{-1}_{*+}{*_\tau}\Big((1{-}e_*^{2iw}){*_\tau}(1{-}e_*^{2iw})^{-1}_{*-}\Big).
$$

\bigskip
Similarly, $\theta_4(w,*)$ is the difference of two inverses of 
$1{+}e_*^{2iw}$
$$
(1{+}e_*^{2iw})^{-1}_{*+}{=}\sum_{n=0}^{\infty}(-1)^ne_*^{2niw},\quad
(1{+}e_*^{2iw})^{-1}_{*-}{=}-\sum_{n=1}^{\infty}(-1)^ne_*^{-2niw}.
$$
Note that
$2e_*^{iw}{*}\sum_{n\geq 0}(-1)^ne_*^{2inw}$ and 
$2e_*^{-iw}{*}\sum_{n\geq 0}(-1)^ne_*^{-2inw}$ are 
 ${*}$-inverses of 
$\frac{1}{2}(e_*^{iw}{+}e_*^{-iw})$. We denote 
these by 
$$
(\cos_*w)^{-1}_{*+}, \quad  (\cos_*w)^{-1}_{*-}.
$$
Then we see  
$$
2i\theta_1(w,*){=}(\cos_*w)^{-1}_{*+}-(\cos_*w)^{-1}_{*-}.
$$

\bigskip

Similarly, 
$2ie_*^{-iw}{*}\sum_{n\geq 0}e_*^{-2inw}$, 
$-2ie_*^{iw}{*}\sum_{n\geq 0}e_*^{2inw}$ are 
both ${*}$-inverses of $\frac{1}{2i}(e_*^{iw}{-}e_*^{-iw})$. We denote 
these by 
$$
(\sin_*w)^{-1}_{*+}, \quad  (\sin_*w)^{-1}_{*-}.
$$
Then we see  
$$
2i\theta_2(w,*){=}(\sin_*w)^{-1}_{*+}-(\sin_*w)^{-1}_{*-}.
$$
Every $\theta_i(w,*)$ is written as the  
difference of two different inverses of suitable linear combinations of 
$*$-exponential functions. 

\medskip
As the associativity fails, $\theta_i(w,{*})$ may be viewed as associators. But, these are
not elements of non-associative {\it algebra}, for the product  
$(\sum_{n=0}^{\infty}e_*^{niw}){*}(\sum_{-\infty}^{-1}e_*^{niw})$ is
not defined as it diverges apparently.

\subsubsection{Star-delta functions}

Next, we note a similar phenomenon 
as above for the generator of the algebra: 

\begin{prop}\label{geninverse}
If ${\rm{Re}}\,\tau{>}0$, then for every 
$a{\in}{\mathbb C}$, the integrals 
$$
i\!\int_{-\infty}^0{:}e_*^{it(a+w)}{:}_\tau dt, \quad  
{-}i\!\int_{0}^{\infty}{:}e_*^{it(a+w)}{:}_\tau dt
$$ 
converge in $H\!ol(\mathbb C)$ to give inverses of $a{+}w$. 
If $a=-i\alpha$, then 
$$
\int_{-\infty}^0{:}e_*^{t(\alpha+iw)}{:}_\tau dt, \quad  
{-}\int_{0}^{\infty}{:}e_*^{t(\alpha+iw)}{:}_\tau dt
$$
converge in $H\!ol(\mathbb C)$ to give inverses of 
$\alpha{+}iw$.
\end{prop}

\noindent
{\bf Proof}\,\,\, For the same reason as 
in the case of $\theta_3$, we get  
the convergence. To confirm these give inverses of $a+w$, 
we compute by the continuity of $i(a{+}w){*}$ as follows: 
$$
(a{+}w){*}i\int_{-\infty}^0e_*^{it(a{+}w)}dt
\,{=}\int_{-\infty}^0i(a{+}w){*}e_*^{it(a{+}w)}dt
\,{=}\int_{-\infty}^0\frac{d}{dt}e_*^{it(a{+}w)}dt{=}1.
$$
The others are obtained similarly. \hfill$\Box$

\bigskip
Denote these inverses by 
$$
{:}(a{+}w)_{*+}^{-1}{:}_{\tau}\,
{=}i\int_{-\infty}^0{:}e_*^{it(a+w)}{:}_{\tau}dt,\quad  
{:}(a{+}w)_{*-}^{-1}{:}_{\tau}\,
{=}{-}\!i\int_{0}^{\infty}{:}e_*^{it(a+w)}{:}_{\tau}dt, 
\quad {\rm{Re}}\,\tau >0. 
$$
$$
{:}(\alpha{+}iw)_{*+}^{-1}{:}_{\tau}\,
{=}\int_{-\infty}^0{:}e_*^{t(\alpha+iw)}{:}_{\tau}dt,\quad  
{:}(\alpha{+}iw)_{*-}^{-1}{:}_{\tau}\,
{=}{-}\!\int_{0}^{\infty}{:}e_*^{t(\alpha+iw)}{:}_{\tau}dt, 
\quad {\rm{Re}}\,\tau >0. 
$$
The definitions of 
${:}(-i\alpha{+}w)_{*+}^{-1}{:}_{\tau}$ and 
${:}(\alpha{+}iw)_{*+}^{-1}{:}_{\tau}$ are the same, since these have the 
same $\tau$-expressions. 
As a result, we have the identity 
$$
(-i\alpha{+}w)_{*+}^{-1}=(-i)^{-1}(\alpha+iw)^{-1}_{*+}
=i(\alpha{+}iw)_{*+}^{-1}.
$$ 
But this is not a special case of 
$\frac{e^{i\theta}}{e^{i\theta}(a+w)}=\frac{1}{a+w}$,
which is trivial in ordinary calculus.
Such an identity cannot be applied in the case 
where $e^{i\theta}$ causes a rotation  
of the path of integration. (See \S\,\ref{rotation}.) 

\medskip
The difference of these two inverses is 
\begin{equation}\label{deltafunc}
(a{+}w)_{*+}^{-1}{-}(a{+}w)_{*-}^{-1}{=}
{i}\int_{-\infty}^{\infty}e_*^{it(a+w)}dt,\quad {\rm{Re}}\,\tau >0.
\end{equation}
The right hand side may be viewed as a delta function 
in the world of $*$-functions. 

We define the $*$-delta function by  
\begin{equation}\label{stardelta}
\delta_*(a{+}w)=\frac{1}{2\pi}\int_{-\infty}^{\infty}e_*^{it(a+w)}dt\quad {\rm{Re}}\,\tau >0.
\end{equation}
It is easy to see that $(a{+}w){*}\delta_*(a{+}w)=0$.  
In contrast, an element $e_*^{2iw}$ cannot be a zero-divisor:
\begin{prop}\label{unique}
${:}e_*^{2iw}{:}_{\tau}{*_\tau}f(w)=e^{-\tau}e^{2iw}f(w{+}i\tau)$. Hence 
a real analytic function $f(w)$ satisfying ${:}e_*^{2iw}{:}_{\tau}{*_\tau}f(w)=0$ must be $0$.
\end{prop}

In ordinary calculus, 
$\int_{-\infty}^{\infty}e^{it(a+x)}dt{=}2\pi\delta(a{+}x)$ is 
not a function but a distribution. On the other hand,
in the world of $*$-functions, 
the $\tau$-expression ${:}\delta_*(a{+}w){:}_\tau$  
of $\delta_*(a{+}w)$ is an entire function: 
\begin{equation}\label{eq:Fouriertrsf}
{:}\delta_*(a{+}w){:}_\tau=
\frac{1}{2\pi}\int_{-\infty}^{\infty}e^{-\frac{1}{4}t^2\tau}e^{it(a{+}w)}dt{=}
\frac{1}{\sqrt{\pi\tau}}e^{-\frac{1}{\tau}(a{+}w)^2}, \quad 
{\rm{Re}}\,\tau >0, \quad a{\in}{\mathbb C}
\end{equation}
This is obtained directly as follows: Set 
$$
-\frac{1}{4}t^2\tau{+}it(a{+}w){=}
-\frac{1}{4}\tau(t{-}\frac{2}{\tau}(a{+}w))^2
{-}\frac{1}{\tau}(a{+}w)^2.
$$
Then we have 
$$
\int_{-\infty}^{\infty}e^{-\frac{1}{4}t^2\tau}e^{it(a{+}w)}dt{=}
\int_{-\infty}^{\infty}
e^{-\frac{\tau}{4}(t{-}\frac{2i}{\tau}(a{+}w))^2}dt
e^{{-}\frac{1}{\tau}(a{+}w)^2}.
$$
Note now that the integral 
$\int_{-\infty}^{\infty}e^{-\frac{\tau}{4}(t{-}\alpha)^2}dt$
converges for every $\alpha\in{\mathbb C}$, and 
integration by parts gives 
$$
\frac{d}{d\alpha}
\int_{-\infty}^{\infty}e^{-\frac{\tau}{4}(t{-}\alpha)^2}dt
=-\int_{-\infty}^{\infty}\frac{d}{dt}
e^{-\frac{\tau}{4}(t{-}\alpha)^2}dt=0. 
$$
Thus, 
$\int_{-\infty}^{\infty}
e^{-\frac{\tau}{4}(t{-}\frac{2}{\tau}(a{+}w))^2}dt$ is 
obtained by the value at $a+w=0$, as  
$\int_{-\infty}^{\infty}
e^{-\frac{\tau}{4}t^2}dt=
2\int_{0}^{\infty}e^{-\frac{\tau}{4}t^2}dt$.

The world of $*$-functions is a deformed version of  
ordinary world, and 
as a result one can see what is not able to see in the ordinary
world. 
Note that there are many inverses, for 
$(a{+}w)_{*+}^{-1}{+}Ci\delta_*(a{+}w)$ 
are all inverses of $a{+}w$ for any $C$.

\subsubsection{Jacobi's imaginary transformations}

By formula \eqref{eq:Fouriertrsf}, we see  
the following series 
$$
\begin{aligned}
&\tilde\theta_1(w ,*){=}
\sum_n(-1)^n\delta_*(w {+}\frac{\pi}{2}{+}\pi n), \quad 
\tilde\theta_2(w ,*){=}\sum_n(-1)^n\delta_*(w{+}\pi n)\\
&\tilde\theta_3(w ,*){=}\sum_n \delta_*(w {+}\pi n), 
\qquad\qquad\quad
\tilde\theta_4(w ,*){=} 
\sum_n \delta_*(w{+}\frac{\pi}{2}{+}\pi n),
\end{aligned}
$$
converge in the $\tau$-expression for ${\rm{Re}}\,\tau>0$. 
These may be viewed as $\pi$-periodic/$\pi$-alternating periodic 
${*}$-delta functions on ${\mathbb R}$. 

As $e^{2\pi in}{=}1$, we have identities 
$$
e_*^{2iw }{*}\tilde\theta_i(w ,*){=}\tilde\theta_i(w ,*), \quad 
(i=2,3),\quad 
e_*^{2iw }{*}\tilde\theta_i(w ,*){=}{-}\tilde\theta_i(w ,*),\quad 
(i=1,4). 
$$
By a slight modification of Proposition\ref{easy00}, we have 
$$
\theta_i(w ,*)=\alpha_i\tilde\theta_i(w ,*), \quad \alpha_i\in{\mathbb C}.
$$
Note that $\alpha_i$ does not depend on the expression parameter $\tau$. 
Taking the $\tau$-expressions of both sides at $\tau{=}\pi$ and  
setting $w{=}0$, we have $\alpha_i=1/2$.

\begin{prop}
$\theta_i(w ,*)=\frac{1}{2}\tilde\theta_i(w ,*)$ for 
$i=1\sim 4$. 

The Jacobi's imaginary transformation is given by 
taking the $\tau$-expression of these identities.
\end{prop}

This may be proved directly by the following manner: Since  
$f(t){=}\sum_ne_*^{2(n+t)iw }$ is periodic function of period $1$, 
Fourier expansion formula gives 
$$
f(t){=}\sum_m \int_0^{1}f(s)e^{-2\pi ims}ds e^{2\pi imt}, \quad 
\theta_3(w,*){=}f(0){=}
\sum_m \int_0^{1}(\sum_ne_*^{2(n+s)iw })e^{-2\pi ims}ds.
$$  
Since $e^{-2\pi ims}=e^{-2\pi im(s+n))}$, we have 
\begin{equation*}\label{thetarel}
\begin{aligned}
f(0){=}&\sum_m \int_0^{1}
(\sum_ne_*^{2(n+s)iw }e^{-2(n+s)i\pi m})ds \\
=&
\sum_m \int_{-\infty}^{\infty}e_*^{2si(w +\pi m)}ds
{=}
\frac{1}{2}\sum_m \delta_*(w +\pi m).
\end{aligned}
\end{equation*}

\medskip 
For instance, we have  
\begin{equation}\label{thetarel}
\begin{aligned}
&\theta_3(w,{\tau})
{=}\frac{2\pi}{2}{:}\sum_n \delta_*(w{+}\pi n){:}_{\tau}
= \sqrt{\frac{\pi}{\tau}}
  \sum_n e^{-\frac{1}{\tau}(w{+}\pi n)^2}\\
&{=}
\sqrt{\frac{\pi}{\tau}}
e^{-\frac{1}{\tau}w ^2}
\sum_n e^{-\pi^2n^2\tau^{-1}-2\pi n\tau^{-1}w}
=
\sqrt{\frac{\pi}{\tau}}
e^{-\frac{1}{\tau}w ^2}
\theta_3(\frac{\pi w}{i\tau},\frac{\pi^2}{\tau})
\end{aligned}
\end{equation}
This is remarkable since a relation of two different 
expressions (viewpoints) are explicitly given. 

In particular, Jacobi's theta relation is 
obtained by setting $w=0$ in \eqref{thetarel}:
\begin{equation}\label{Jacobireleq}
\theta_3(0,{\tau})=
\sqrt{\frac{\pi}{\tau}}\theta_3(0,\frac{\pi^2}{\tau}).
\end{equation}
This will be used  
to obtain the functional identities of the $*$-zeta function 
in the forthcoming paper.

\subsubsection{The boundary of expression parameters}
\label{rotation}

First of all, recall that Fabry-P{\'{o}}lya's gap theorem gives:
\begin{thm}\label{gap}
For a series $\sum_{n=0}^{\infty}a_nq^{n^2}$ suppose 
$a_n{\geq}0$ and the radius of convergence is 1, e.g. the case 
$\limsup_n|a_n|=1$. 
Then  the circle $|q|{=}1$ is the natural boundary.
\end{thm}

Fabry-P{\'{o}}lya's gap Theorem\,\ref{gap} 
shows that the expression parameters 
of $*$-functions such as Jacobi's theta functions  
are strictly limited in the right half plane, that is, complex 
rotations of the domain is forbidden.

\subsubsection{Integration vs discrete summation}
For a while, the expression parameter is restricted in 
the domain ${\rm{Re}}\,\tau<0$. 
The difference between the integral $\int_{-\infty}^0e_*^{tw }dt$ 
and the discrete sum  
$\sum_{n=0}^{\infty}e_*^{nw }$ appears in the following 
logarithmic derivative 
$(\frac{\partial}{\partial\beta}\log)f(\beta)
=f'(\beta){*}f(\beta)^{-1}$:

\begin{equation}
  \label{eq:differ}
\sum_{k=1}^{\infty}(\frac{\partial}{\partial\beta}\log)
\int_{-\infty}^0e_*^{t\beta kw }dt,\qquad 
\sum_{k=1}^{\infty}(\frac{\partial}{\partial\beta}\log)
\sum_{n=1}^{\infty}e_*^{n\beta kw }.   
\end{equation}
Since integration by parts gives 
$$
\frac{\partial}{\partial\beta}\int_{-\infty}^0e_*^{t\beta kw }dt
=\beta^{-1}\int_{-\infty}^0t(k\beta w ){*}e_*^{tk\beta w }dt=
{-}\beta^{-1}\int_{-\infty}^0e_*^{t\beta kw }dt,
$$
the first term of \eqref{eq:differ} is 
$\sum_{k=1}^{\infty}\beta^{-1}$ and there 
is no way to avoid divergence. However, 
the second term of \eqref{eq:differ} is rewritten as 
$\sum_{k=1}^{\infty}\big(\sum_{n=1}^{\infty}e_*^{n\beta k}\big){*}kw$
by using the identity 
$$
\frac{\partial}{\partial\beta}
\sum_{n=1}^{\infty}e_*^{n\beta kw }=
kw {*}\sum_{n=1}^{\infty}ne_*^{n\beta kw }=
kw {*}\big(\sum_{n=1}^{\infty}e_*^{n\beta kw }\big){*}
\big(\sum_{n=1}^{\infty}e_*^{n\beta kw }\big)
$$
which is easily proved.  
The $\tau$-expression of 
$\sum_{k=1}^{\infty}\big(\sum_{n=1}^{\infty}e_*^{n\beta k}\big){*}kw$ 
 converges absolutely when ${\rm{Re}}\,\tau{<}0$ by the next Lemma:
\begin{lem}
\label{hyouka}
If 
${\rm{Re}}\,\tau{<}0$, then the $\tau$-expression of 
$\sum_{k=1}^{\infty}\sum_{n=1}^{\infty}ke_*^{n\beta kw }$
converges absolutely. 
\end{lem}

\noindent
{\bf Proof}\,\, The $\tau$-expression is given by 
${:}e_*^{n\beta kw }{:}_{\tau}{=}
e^{nk\beta w {+}\frac{1}{4}(nk\beta)^2\tau}$.  Rewriting this as 
$$
nk\beta w {+}\frac{1}{4}(nk\beta)^2\tau{=}
\frac{\tau}{4}(\beta nk{+}\frac{2}{\tau}w )^2
{-}\frac{1}{\tau}w ^2
$$
and assuming  ${\rm{Re}}(\tau\beta^2){<}0$, we only have 
to show the 
convergence of 
$\sum_{k,n=1}^{\infty}k
e^{\frac{\tau}{4}(\beta nk{+}\frac{2}{\tau}w )^2}$.
Restricting $w $ to any compact subset, it is enough to show 
the absolute convergence of   
$\sum_{k,n=1}^{\infty}k e^{\frac{\tau}{4}\beta^2(nk{+}K)^2}$ 
for every $K>0$.

For $n,k{\geq}1$, 
$(nk{+}K)^2{\leq}(n^2{+}1)^2{+}(k^2+1)^2{+}K^2$ gives 
$$
\sum_{k,n=1}^{\infty}k e^{\frac{\tau}{4}\beta^2(nk{+}K)^2}
\leq 
e^{\frac{1}{4}\tau\beta^2K^2}
\sum_{n=1}^{\infty}e^{\frac{1}{4}\tau\beta^2(n^2{+}1)^2}
\sum_{k=1}^{\infty}ke^{\frac{1}{4}\tau\beta^2(k^2{+}1)^2}.
$$
This gives convergence. \hfill $\Box$

\section{Calculus of inverses}

In this section, we give several ways of treatment of $*$-product of 
inverse elements. First of all, we note 
that the elementary method of constant variation gives many 
inverses of a single element. Let $a\in{\mathbb C}$. 
By the product formula, 
$(a{+}w)*$ is viewed as a linear operator of 
$H\!ol({\mathbb C})$ into itself. Suppose $\tau\not=0$. 
Then, $(a{+}w)*_{\tau}f(w)=0$ is a differential equation 
$$
(a{+}w)f(w){+}\frac{\tau}{2}\partial_{w}f(w)=0.
$$
Solving this, we have   
\begin{equation}\label{deltaexp}
(a{+}w)*_{\tau}Ce^{-\frac{1}{\tau}(a{+}w)^2}= 
Ce^{-\frac{1}{\tau}(a{+}w)^2}{*_\tau}(a{+}w)=0.
\end{equation}
The method of constant variation gives a function 
$g_a(w)$ such that 
$$
(a{+}w){*_{\tau}}g_a(w)=g_a(w){*_\tau}(a{+}w)=1.
$$  
Thus, we have   
\begin{equation}
g_{a}(w)=
\frac{2}{\tau}
\int_0^{1}e^{\frac{1}{\tau}((a{+}wt)^2{-}(a{+}w)^2)}wdt{+}C
e^{-\frac{1}{\tau}(a{+}w)^2}.
\end{equation}
This breaks associativity 
$$
(e^{-\frac{1}{\tau}(a{+}w)^2}{*_\tau}(a{+}w))*_{\tau}g_a(w)\not=
e^{-\frac{1}{\tau}(a{+}w)^2}{*_\tau}((a{+}w)*_{\tau}g_a(w)).
$$
The method of constant variation is a strange method from a viewpoint
of operators. 

\medskip 
Note that if $a\not=b$, then 4 elements with independent
$\pm$-sign 
$$
\frac{1}{b{-}a}(\big((a{+}w)_{*\pm}^{-1}-(b{+}w)_{*\pm}^{-1}\big))
$$
give respectively ${*}$-inverses of $(a{+}w){*}(b{+}w)$.
Thus, we define ${*}$-inverse with independent $\pm$-sign by 
\begin{equation}\label{defprodinv}
(a{+}w)_{*\pm}^{-1}{*}(b{+}w)_{\pm}^{-1}=
\frac{1}{b{-}a}(\big((a{+}w)_{*\pm}^{-1}-(b{+}w)_{*\pm}^{-1}\big)).
\end{equation}
By providing $a\not=b$, a direct calculation of ${*}$-product by 
\eqref{defprodinv} gives  
$$
\Big((a{+}w)^{-1}_{*+}{-}(a{+}w)^{-1}_{*-}\Big){*}
\Big((b{+}w)^{-1}_{*+}{-}(b{+}w)^{-1}_{*-}\Big)=0, \quad 
a,b\in {\mathbb C}, \,\,a\not=b.
$$

\begin{thm}\label{prodtwoinv}
Suppose $a{+}w$ has two different $*$-inverses 
$(a{+}w)^{-1}_{*+}$ and $(a{+}w)^{-1}_{*-}$, then 
$$
(a{+}w){*}((a{+}w)^{-1}_{*+}{-}(a{+}w)^{-1}_{*-})=0.
$$ 
If $b{+}w$ has also two different $*$-inverses, then by providing $a\not=b$,
the ${*}$-product 
$$
\Big((a{+}w)^{-1}_{*+}{-}(a{+}w)^{-1}_{*-}\Big){*}
\Big((b{+}w)^{-1}_{*+}{-}(b{+}w)^{-1}_{*-}\Big)=0, \quad 
a\not=b.
$$
\end{thm}

In the later section, we show the formula 
$$
\delta_*(a{+}w){*}\delta_*(b{+}w)=\delta(a{-}b)\delta_*(b{+}w).
$$

\medskip
Existence of two different inverses means that there are
two ways to define {\it logarithmic derivatives}: 
$$
\frac{d}{d\alpha}\log_{*+}(\alpha{+}iw)=(\alpha{+}iw)_{*+}^{-1},\quad 
\frac{d}{d\alpha}\log_{*-}(\alpha{+}iw)=(\alpha{+}iw)_{*-}^{-1},
$$
while $\frac{d}{d\alpha}\log_{*\pm}e_*^{(\alpha{+}iw)}=\alpha{+}iw$.

\medskip
Note also that if ${\rm{Re}}\,\tau>0$, then $(z{+}w)_{*\pm}^{-1}$ is
an entire function with respect to $z$, and hence the integral on any 
closed path $C$  
$$
\int_C f(z)(z{+}w)_{*\pm}^{-1}dz =0
$$ 
for every entire function $f(z)$. Hence, we cannot use Cauchy's
integral formula.   

\subsection{Resolvent calculus} 

Products of inverses are given for $k, \ell \geq 0$ by 
\begin{equation}\label{resolvent}
(\alpha{+}w)_{*+}^{-k{-}1}{*}(\beta{+}w)_{*+}^{-\ell{-}1}
= (-1)^{k+\ell}\frac{1}{k!{\ell}!}\frac{d^{k{+}\ell}}{d\alpha^kd\beta^{\ell}}
\frac{1}{\beta{-}\alpha}
\big((\alpha{+}w)_{*+}^{-1}-(\beta{+}w)_{*+}^{-1}\big).
\end{equation}
Note that $(\alpha{+}w)_{*+}^{-1}{*}(\alpha{+}w)_{*+}^{-1}$ is
defined by this formula. 
Such calculations of $*$-inverses will be referred as  
the {\bf resolvent calculus}.

We now make $*$-inverses of $p_*(w)$ by the partial fraction
decomposition as follows: Setting $X=w$ for simplicity and 
regarding $p(X)$ is an ordinary polynomial, we see  
$$
\frac{1}{p(X)}=\sum_{k=1}^R\frac{c_k}{(a_k{-}X)^{m_k}},\quad
c_k\in \mathbb C.
$$ 
Since ${\mathbb C}[w_*]$ is isomorphic to the ordinary polynomial
ring ${\mathbb C}[X]$, we see 
\begin{equation}\label{polyinvers}
p_*(w)_{*\pm}^{-1}=\sum_{k=1}^Rc_k(a_k{-}w)_{*\pm}^{-m_k}
\end{equation}
is well-defined by \eqref{resolvent} to give $*$-inverses of
$p_*(w)$. That is, every ${*}$-polynomial has $*$-inverses.

\medskip
In these computations, the following proposition is very useful:
\begin{prop}
Suppose $w$ is restricted to a compact domain.  
If ${\rm{Re}}\,\tau{>}0$, then   
${:}(z{+}w)_{*\pm}^{-1}{:}_{\tau}$ are both 
rapidly decreasing functions w.r.t. $z$ on the closed 
lower/upper complex half-plane.  
\end{prop} 

\noindent
{\bf Proof}\,\,  
Set $z{=}x{+}iy$. We will show that 
 $z^k\partial_z^{\ell}{:}(z{+}w)_{*+}^{-1}{:}_{\tau}$ 
are bounded on the closed lower half-plane. 
$$
z^k\partial_z^{\ell}{:}(z{+}w)_{*+}^{-1}{:}_{\tau}= 
i\int_{-\infty}^0(it)^{\ell}((-i\partial_t)^ke^{itz})
e^{-\frac{1}{4}\tau t^2}e^{itw}dt.
$$
Integration by parts gives that there is a polynomial $p(t, w)$ 
depending on $k,\ell$ such that 
$$
z^k\partial_z^{\ell}{:}(z{+}w)_{*+}^{-1}{:}_{\tau}=
i\int_{-\infty}^0p(t,w)e^{itz}{:}e_*^{itw}{:}_{\tau}dt. 
$$ 
If $z$ is in the lower half-plane then $|e^{itz}|{\leq}1$. The
boundedness follows. 
The other case is proved similarly. \hfill $\Box$. 

Hence we have the same result as in \eqref{eq:Fouriertrsf}:  
\begin{cor}\label{corcor00}
Suppose ${\rm{Re}}\,\tau{>}0$. Then 
$\delta_*(z{+}w)$ is rapidly decreasing on the real line.
\end{cor}

Since $(-z{+}w)_{*-}^{-1}=-(z{-}w)_{*-}^{-1}$ is rapidly 
decreasing w.r.t. $z$ on the right half-plane, we often 
write these as follows: 
$$
(z{-}w)_{*-}^{-1}=i\int_{-\infty}^0e_{*}^{it(z{-}w)}dt, 
$$
or similarly  
$$
(z{+}w)_{*+}^{-1}=-i\int_0^{\infty}e_{*}^{-it(z{+}w)}dt. 
$$

\bigskip
 
Note that although 
$(\alpha{+}w)_{*+}^{-1}{*}(\alpha{+}w)_{*-}^{-1}$ diverges,  
the product 
$(\alpha{+}w)_{*+}^{-k}{*}(\beta{+}w)_{*-}^{-\ell}$ is defined, 
if $\alpha\not=\beta$,  as 
$$
(-1)^{k+\ell}\frac{1}{k!{\ell}!}\frac{d^{k{+}\ell}}{d\alpha^kd\beta^{\ell}}
\frac{1}{\beta{-}\alpha}
\big((\alpha{+}w)_{*+}^{-1}-(\beta{+}w)_{*-}^{-1}\big).
$$

To compute the $*$-product of mixed elements, we first replace 
$(\alpha{+}w)_{*-}^{-1}$ by 
$(\alpha{+}w)_{*+}^{-1}{-}\delta_*(\alpha{+}w)$, 
and then we have only 
to note the following a little strange identities: 
\begin{equation}\label{sekidelta}
\delta_*(\alpha{+}w){*}\delta_*(\beta{+}w)=0, \quad 
\alpha\not=\beta, \quad {\text{cf.}} \eqref{sekidlt}
\end{equation}
\begin{equation}
(\alpha{+}w)_{*+}^{-1}{*}\delta_*(\beta{+}w)=
\frac{1}{\alpha{-}\beta}\delta_*(\beta{+}w)\quad (\alpha\not=\beta). 
\end{equation}

Since $\delta_*(a{+}w){*}w=-a\delta_*(a{+}w)=w{*}\delta_*(a{+}w)$, 
\eqref{sekidelta} barely protect the associativity 
$$
(\delta_*(a{+}w){*}w){*}\delta_*(b{+}w)=
\delta_*(a{+}w){*}(w{*}\delta_*(b{+}w)).
$$

\medskip
We prove these formulas directly by integration. In what follows, 
we assume ${\rm{Re}}(a{-}b)>0$, but this is not essential:
$$
\begin{aligned}
(a+w)_{*+}^{-1}{*}(b+w)_{*+}^{-1}&
=-\int_{-\infty}^0\int_{-s'}^{s'}
e^{\frac{1}{2}i\sigma(a-b)}d\sigma 
e^{\frac{1}{2}is'(a+b)}e^{is'w}ds'\\
=&
\frac{1}{b{-}a}\big((a+w)_{*+}^{-1}{-}(b+w)_{*+}^{-1}\big).
\end{aligned}
$$
$$
(a+w)_{*+}^{-1}{*}(b+w)_{*-}^{-1}=
(a+w)_{*+}^{-1}{*}
\big((b+w)_{*+}^{-1}{-}\delta_*(b{+}w)\big)
$$
$$
(a{+}w)_{*+}^{-1}{*}\delta_*(b{+}w)=
\int_{-\infty}^0\int_{\mathbb R}
e^{i(a-b)t'}e^{is'b}e_*^{is'w}dt'ds'
=\frac{1}{a-b}\delta_*(b{+}w).
$$

\subsection{The case of inverses by discrete summations} 

Suppose ${\rm{Re}}\,\tau>0$, then for every $a\in{\mathbb C}$, the
bilateral geometric series 
$\sum_{n\in\mathbb Z}{:}e_*^{2in(w{+}a)}{:}_{\tau}$ converges. Denote
this  by $\theta_3(w{+}a,*)$. We see easily that 
$$
\theta_3(w{+}a,*) =
(1{-}e^{2ia}e_*^{2iw})^{-1}_{*+}{-}(1{-}e^{2ia}e_*^{2iw})^{-1}_{*-}.
$$
For another $b\in{\mathbb C}$, we make $\theta_3(w{+}b,*)$ and 
$$
\theta_3(w{+}b,*)=
(1{-}e^{2ib}e_*^{2iw})^{-1}_{*+}{-}(1{-}e^{2ib}e_*^{2iw})^{-1}_{*-}.
$$
As $e^{2ia}$, $e^{2ib}$ are scalars, we see that $4$ independent elements
$$
\frac{1}{e^{2ia}{-}e^{2ib}}
\Big((1{-}e^{2ia}e_*^{2iw})^{-1}_{*\pm}{-}(1{-}e^{2ib}e_*^{2iw})^{-1}_{*\pm}\Big)\quad
(\text{independent $\pm$ sign})
$$
gives respectively inverses of
$(1{-}e^{2ia}e_*^{2iw}){*}(1{-}e^{2ib}e_*^{2iw})$. Hence, we define
the $*$-product of inverses by such resolvent calculations, that is, we
define 
$$
(1{-}e^{2ia}e_*^{2iw})^{-1}_{*\pm}{*}(1{-}e^{2ib}e_*^{2iw})^{-1}_{*\pm}=
\frac{1}{e^{2ia}{-}e^{2ib}}
\Big((1{-}e^{2ia}e_*^{2iw})^{-1}_{*\pm}{-}(1{-}e^{2ib}e_*^{2iw})^{-1}_{*\pm}\Big),
$$
although there are many choice of inverses.

\begin{thm}\label{vanishingthm}
If $a\not=b$, then under such a definition of $*$-product of inverses,
we have $\theta_3(w{+}a,*){*}\theta_3(w{+}b,*)=0$. In particular 
$\theta_3(w,*){*}\theta_4(w,*)=0$ by noting 
$\theta_3(w{+}\frac{\pi}{2},*)=\theta_4(w,*)$.  It follows by
differentiating  
$$
(\partial_w\theta_3(w,*)){*}\theta_4(w,*){+}\theta_3(w,*){*}\partial_w\theta_4(w,*)=0.
$$ 
\end{thm}

\noindent
{\bf Proof}\,\,Using 
$\theta_3(w{+}a,*) =
(1{-}e^{2ia}e_*^{2iw})^{-1}_{*+}{-}(1{-}e^{2ia}e_*^{2iw})^{-1}_{*-}$, 
we compute the ${*}$-product 
$$
\Big((1{-}e^{2ia}e_*^{2iw})^{-1}_{*+}{-}(1{-}e^{2ia}e_*^{2iw})^{-1}_{*-}\Big){*}
\Big((1{-}e^{2ib}e_*^{2iw})^{-1}_{*+}{-}(1{-}e^{2ib}e_*^{2iw})^{-1}_{*-}\Big)
$$
by the resolvent calculus. By multiplying $e^{2ia}{-}e^{2ib}$, we see 
$$
\begin{aligned}
   (1{-}e^{2ia}e_*^{2iw})^{-1}_{*+}{-}&(1{-}e^{2ib}e_*^{2iw})^{-1}_{*+}\\
{-}(1{-}e^{2ia}e_*^{2iw})^{-1}_{*+}{+}&(1{-}e^{2ib}e_*^{2iw})^{-1}_{*-}\\
{-}(1{-}e^{2ia}e_*^{2iw})^{-1}_{*-}{+}&(1{-}e^{2ib}e_*^{2iw})^{-1}_{*+}\\
{+}(1{-}e^{2ia}e_*^{2iw})^{-1}_{*-}{-}&(1{-}e^{2ib}e_*^{2iw})^{-1}_{*-}.
\end{aligned}
$$
We see all terms are cancelled out.
\hfill $\Box$

The trivial identity gives 
$e_*^{2iw}{*}\theta_3(w{+}a,*)=\theta_3(w{+}a,*){*}e_*^{2iw}=e^{-2ia}\theta_3(w{+}a,*)$ and the
vanishing Theorem\,\ref{vanishingthm} barely protects the associativity:
$$
\Big(\theta_3(w{+}a,*){*}e_*^{2iw}\Big){*}\theta_3(w{+}b,*)=
\theta_3(w{+}a,*){*}\Big(e_*^{2iw}{*}\theta_3(w{+}b,*)\Big),\quad a\not=b.
$$

\bigskip
We keep the expression parameter $\tau$ so that  ${\rm{Re}}\,\tau>0$.
Let ${\mathbb C}[e_*^{iw}]$ be the polynomial ring of $e_*^{iw}$. 
For every monic polynomial $p(X)$ (with $1$ as the coefficient of the
highest degree), we set by unique factorization theorem
$$
p(e_*^{iw})=\prod_{k=1}^N(a_k{-}e_*^{iw})^{\ell_k}.
$$
For a special polynomial $X^n$, we use the notation $I^{(n)}(e_*^{iw})=e_*^{niw}$.
Note that 
\begin{equation}\label{highinv}
(1{-}X)_{*\pm}^{-n-1}=\frac{1}{(n-1)!}\partial_X^n(1{-}X)_{*\pm}^{-1}
\end{equation}
are well-defined to give inverses of $(1{-}X)_*^{n+1}$.

We now make $*$-inverses of $p(e_*^{iw})$ by the partial fraction
decomposition as follows: Setting $X=e_*^{iw}$ for simplicity and 
regarding $p(X)$ is an ordinary polynomial, we see  
$$
\frac{1}{p(X)}=\sum_{k=1}^R\frac{c_k}{(a_k{-}X)^{m_k}},\quad
c_k\in \mathbb C.
$$ 
Since ${\mathbb C}[e_*^{iw}]$ is isomorphic to the ordinary polynomial
ring ${\mathbb C}[X]$, we see 
$$
p(e_*^{iw})_{*\pm}^{-1}=\sum_{k=1}^Rc_k(a_k{-}e_*^{iw})_{*\pm}^{-m_k}
$$
is well-defined by \eqref{highinv} to give $*$-inverses of
$p(e_*^{iw})$, where we have to note that 
$$
(e_*^{iw})_{*+}^{-1}=e_*^{-iw}=(e_*^{iw})_{*-}^{-1}, \quad\text{(cf.Proposition\,\ref{unique})} 
$$
while in the continuous case, Proposition\,\ref{geninverse} shows that 
$w$ has two different inverses $w_{*+}^{-1}$, $w_{*-}^{-1}$. 

\medskip
As in the case of $*$-theta functions we define 
$$
{\hat P}_*(e_*^{iw})=p(e_*^{iw})_{*+}^{-1}{-}p(e_*^{iw})_{*-}^{-1},
\quad {\hat{I}}_*^{(n)}(e_*^{iw})=0.
$$
This is $2\pi$-periodic, and providing $p(0)=1$ and 
setting $p(e_*^{iw})=1{-}q_*(e_*^{iw})$, we see 
$$
q_*(e_*^{iw}){*}{\hat P}_*(e_*^{iw})={\hat P}_*(e_*^{iw}).
$$
As a Corollary of Fabry-P{\'{o}}lya's gap Theorem\,\ref{gap} we see 
the following: 
\begin{cor}\label{fabry}
For every polynomial $p(X)$, ${:}p(e_{*\pm}^{iw})^{-1}{:}_{\tau}$ are
holomorphic w.r.t. $\tau$ on the right half-plane ${\rm{Re}}\,\tau >0$, and  
${\rm{Re}}\,\tau=0$ is the natural boundary except the case $p(X)=aX^n$. 
\end{cor}

\bigskip 
If another polynomial $q(X)$ is relatively prime to $p(X)$, then
there are polynomials $a(X)$, $b(X)$ such that 
$a(X)p(X){+}b(X)q(X)=1$. Hence, we have that 4-independent elements
$$
b(e_*^{iw}){*}p(e_*^{iw})_{*\pm}^{-1}{+}a(e_*^{iw}){*}q(e_*^{iw})_{*\pm}^{-1}
$$ 
are inverses of $p(e_*^{iw}){*}q(e_*^{iw})$ respectively. Hence, we
define the $*$-products of these inverses by 
$$
p(e_*^{iw})_{*\pm}^{-1}{*}q(e_*^{iw})_{*\pm}^{-1}=
b(e_*^{iw})p(e_*^{iw})_{\pm}^{-1}{+}a(e_*^{iw})q(e_*^{iw})_{\pm}^{-1}
$$
by using resolvent calculations. We have then a little strange theorem 
\begin{thm}\label{strange000}
If $p(X)$, $q(X)$ are relatively prime, that is, the resultant of
$p(X)$ and $q(X)$ does not vanish, 
 then the $*$-product ${\hat P}_*(e_*^{iw}){*}{\hat Q}_*(e_*^{iw})=0$.
\end{thm}

\noindent
{\bf Proof}\,\,is the same as in Theorem\,\ref{vanishingthm}. We
compute as follows: 
$$
\begin{aligned}
\Big(p(e_*^{iw})_{*+}^{-1}{-}p(e_*^{iw})_{*-}^{-1}\Big){*}
\Big(q(e_*^{iw})_{*+}^{-1}{-}q(e_*^{iw})_{*-}^{-1}\Big) =&
\quad \,\,b(e_*^{iw})p(e_*^{iw})_{*+}^{-1}{+}a(e_*^{iw})q(e_*^{iw})_{*+}^{-1}\\
&-b(e_*^{iw})p(e_*^{iw})_{*+}^{-1}{-}a(e_*^{iw})q(e_*^{iw})_{*-}^{-1}\\
&-b(e_*^{iw})p(e_*^{iw})_{*-}^{-1}{-}a(e_*^{iw})q(e_*^{iw})_{*+}^{-1}\\
&+b(e_*^{iw})p(e_*^{iw})_{*-}^{-1}{+}a(e_*^{iw})q(e_*^{iw})_{*-}^{-1}=0.
\end{aligned}
$$
${}$\hfill $\Box$

This sounds strange because the condition for the resultant is an open
condition, while the conclusion looks like a closed condition.

\bigskip
\noindent
{\bf Note}\,\, In conformal field theory, a formal distribution
$\delta(z)=\sum_{n\in\mathbb Z}z^n$ is often used. As we have seen,
this is written as the difference of two inverses 
$\delta(z)=(1{-}z)_{+}^{-1}{-}(1{-}z)_{-}^{-1}$ where 
$$
(1{-}z)_{+}^{-1}=\frac{1}{1{-}z},\quad 
(1{-}z)_{-}^{-1}=-\frac{z^{-1}}{1{-}z^{-1}}. 
$$
We may extend this to define $\delta(az)=\sum_{n\in\mathbb Z} (az)^n$
for $a\in\mathbb C$. Then, we easily see that 
$$
\delta(az)=(1{-}az)_{+}^{-1}{-}(1{-}az)_{-}^{-1}
$$
and if $a\not=b$, then $\delta(az)\delta(bz)=0$ by the same
calculation as in Theorem \,\ref{vanishingthm} below. 

Since the trivial identity $az\delta(az)=\delta(az)$ gives 
$z\delta(az)=a^{-1}\delta(az)$, the associativity holds    
$$
a^{-1}\delta(az)\delta(bz)=(\delta(az)z)\delta(bz)=
\delta(az)(z\delta(bz))=b^{-1}\delta(az)\delta(bz).
$$

\subsubsection{Half-series algebra}

It is well known that 
if a formal power series satisfies $\sum_{n=0}^{\infty}a_nz^n{=}0$, 
then $a_n=0$. 
This is proved by setting $z{=}0$ to get $a_0=0$, and then
taking $\partial_z$ followed by evaluating at $z=0$ gives $a_1{=}0$ and so on.  
Hence this method cannot be applied to formal power series 
$\sum_{n=0}^{\infty}a_ne_*^{niw}$.

We assume ${\rm{Re}}\,\tau>0$ throughout this subsection. 
A formal power series 
$$
z^{\ell}\sum_{n=0}^{\infty}a_nz^n,\quad \ell \in\mathbb Z
$$
 is called a {\it convergent power series}, if 
$\sum_{n=0}^{\infty}a_nz^n$ has a positive radius of convergence.
It is easy to show that if 
$z^{\ell}\sum_{n=0}^{\infty}a_nz^n$ is a convergent power series, then 
$$
f(w)={:}e_*^{\ell iw}{*}\sum_{n=0}^{\infty}a_ne_*^{niw}{:}_{\tau}
$$ 
is an entire function of $w$. Hence if $f(w)=0$, then Proposition\ref{ketugoprodexp22} gives 
$\sum_{n=0}^{\infty}a_ne_*^{niw}{:}_{\tau}=0$, and $a_0=0$ by taking
$w\to i\infty$. Thus, the repeated use of Proposition\ref{ketugoprodexp22} gives 
all $a_n=0$.

Note that the product of two convergent power series is a convergent
power series. 
If 
$
z^{\ell}\sum_{n=0}^{\infty}a_nz^n,\quad (\ell \in \mathbb Z)
$
 is a convergent power series, then its inverse 
$z^{-\ell}(\sum_{n=0}^{\infty}a_nz^n)^{-1}$ obtained by 
the method of indeterminate constants is also a convergent power series. 

We denote by 
${\mathfrak H}_+$ be the space of power series 
${:}e_*^{\ell iw}{*}\sum_{n=0}^{\infty}a_ne_*^{niw}{:}_{\tau}$
made by convergent power series  $z^{\ell}\sum_{n=0}^{\infty}a_nz^n$.
We call ${\mathfrak H}_+$ the half-series algebra. Its fundamental
property is 
\begin{thm}\label{nicefield}
$({\mathfrak H}_+, {*_{\tau}})$ is a topological field consisting of
$2\pi$-periodic entire functions of $w$.  
\end{thm}

\noindent
{\bf Proof}\, is completed by showing the uniqueness of the
inverse. It is reduced to show that 
$$
\sum_{n=0}^{\infty}a_ne_*^{niw}{:}_{\tau}{*_{\tau}}\sum_{k=0}^{\infty}b_ke_*^{kiw}{:}_{\tau}=0
$$
and $a_0\not=0$ gives $\sum_{k=0}^{\infty}b_ke_*^{kiw}{:}_{\tau}=0$. 
The repeated use of Proposition\ref{ketugoprodexp22} gives 
all $b_n=0$. \hfill $\Box$

\subsubsection{Euler numbers}\,\, 
Recall the the generating function of Euler numbers 
$$
\frac{2}{e^{z}{+}e^{-z}}=
\frac{e^z}{1{+}e^{2z}}{+}\frac{e^{-z}}{1{+}e^{-2z}}{=}
\sum_{n=0}^{\infty}E_{2n}\frac{1}{(2n)!}z^{2n},\quad |z|<\pi.
$$ 
The l.h.s.  is a convergent power series obtained by the method of indeterminate constants.
Hence, by Proposition\,\ref{entirelevel} gives 
\begin{equation}\label{Euler}
e_*^{e_*^{iw}}{*}\Big(1{+}\sum_{k=0}^{\infty}{2^ke_*^{kiw}}\frac{1}{k!}\Big)^{-1}
{+}e_*^{-e_*^{iw}}{*}\Big(1{+}\sum_{k=0}^{\infty}{(-2)^ke_*^{kiw}}\frac{1}{k!}\Big)^{-1}=
\sum_{n=0}^{\infty}E_{2n}\frac{1}{(2n)!}e_*^{2niw},
\end{equation}
where $e_*^{\pm e_*^{iw}}=\sum_{\ell=0}^{\infty}\frac{(\pm 1)^{\ell}}{\ell !}
e_*^{\ell iw}$. Note that this identity holds for every expression
$\tau$ such that ${\rm{Re}}\,\tau{>}0$.

\bigskip
On the other hand, by using the formal power series of $(iw)_*^n$, 
we can compute the inverses 
$\Big(1{+}\sum_{k=0}^{\infty}\frac{(2iw)_*^k}{k!}\Big)^{-1}$, 
$\Big(1{+}\sum_{k=0}^{\infty}\frac{(-2iw)_*^k}{k!}\Big)^{-1}$  
by the method of indeterminate constants. Hence, we have also 
\begin{equation}\label{Euler22}
e_*^{iw}{*}\Big(1{+}\sum_{k=0}^{\infty}{(2iw)_*^k}\frac{1}{k!}\Big)^{-1}
{+}e_*^{-iw}{*}\Big(1{+}\sum_{k=0}^{\infty}(-2iw)_*^k\frac{1}{k!}\Big)^{-1}=
\sum_{n=0}^{\infty}E_{2n}\frac{1}{(2n)!}(iw)_*^{2n}.
\end{equation} 
where r.h.s. is a formal power series of $iw$ in the $*$-product.
It is clear that replacing $(iw)_*^k$ by $e_*^{kiw}$ in \eqref{Euler22}
gives  \eqref{Euler}. 

It is very interesting to compare the l.h.s. with  
$e_*^{iw}{*}(1{+}e_*^{2iw})_{*+}^{-1}{+}e_*^{-iw}{*}(1{+}e_*^{-2iw})_{*+}^{-1}$.
The $\tau$-expression of this is an entire function and its 
Taylor expansion is given by 
\begin{equation}
{:}e^{iw}{*}\big(1{+}e_*^{2iw}\big)_{*+}^{-1}
{+}e^{-iw}{*}\big(1{+}e_*^{-2iw}\big)_{*+}^{-1}{:}_{\tau}
=
2\sum_{\ell=0}^{\infty}\sum_{n=0}^{\infty}
(-1)^ne^{-\frac{(2n{+}1)^2}{4}\tau}(2n{+}1)^{2\ell}\frac{1}{(2\ell)!}(iw)^{2\ell}.
\end{equation}
However, this is a $2\pi$-periodic function, while 
$\sum_{n=0}^{\infty}E_{2n}\frac{1}{(2n)!}(iw)_*^{2n}$ is not
$2\pi$-periodic in the $\tau$-expression. Note also that the l.h.s. of \eqref{Euler22} may be
computed as a $2\pi$-periodic function, but the method of indeterminate constants ignores the
periodicity.  
Therefore, we have to rewrite \eqref{Euler22} within
$2\pi$-periodic functions to compare this with  
$e_*^{iw}{*}(1{+}e_*^{2iw})_{*+}^{-1}{+}e_*^{-iw}{*}(1{+}e_*^{-2iw})_{*+}^{-1}$.

\medskip
Note that for every integer $m$, we have 
\begin{equation*}
\begin{aligned}
e_*^{iw}{*}\Big(1{+}\sum_{k=0}^{\infty}{(2i(w{+}2\pi m))_*^k}\frac{1}{k!}\Big)^{-1}
\!\!{+}&e_*^{-iw}{*}\Big(1{+}\sum_{k=0}^{\infty}(-2i(w{+}2\pi
m))_*^k\frac{1}{k!}\Big)^{-1}\\
&{=}
\sum_{n=0}^{\infty}E_{2n}\frac{1}{(2n)!}(i(w{+}2\pi m))_*^{2n}.
\end{aligned}
\end{equation*} 
We will come back again to this problem in the forthcoming paper.

\subsubsection{Bernoulli numbers}\,\, 

Recall the generating function of Bernoulli numbers:
$$
z\Big(\frac{1}{2}{+}\frac{1}{e^z{-}1}\Big){=}
\frac{z}{2}\Big(\frac{1}{e^{z}{-}1}{-}\frac{1}{e^{-z}{-}1}\Big)
{=}\sum_{n=0}^{\infty}B_{2n}\frac{1}{(2n)!}z^{2n}. 
$$
Since $\frac{z}{e^z{-}1}$ and $\frac{-z}{e^{-z}{-}1}$ are 
computed by the method of indeterminate constants as 
$$
(\sum_n\frac{z^n}{(n+1)!})^{-1}
=\sum B_{2n}\frac{1}{(2n)!}z^{2n}{-}\frac{1}{2}z,\quad 
(\sum_n\frac{(-z)^n}{(n+1)!})^{-1}
=\sum B_{2n}\frac{1}{(2n)!}z^{2n}{+}\frac{1}{2}z,
$$
these must hold in the ${*}$-product by replacing $z$ by $e_*^{iw}$. Hence, we have 
\begin{equation}\label{Bern}
\frac{1}{2}\Big(\sum_{n=0}^{\infty}\frac{(e_*^{iw})^n}{(n+1)!}\Big)^{-1}{+}
\frac{1}{2}\Big(\sum_{n=0}^{\infty}\frac{(-e_*^{iw})^n}{(n+1)!}\Big)^{-1}
=\sum_{n=0}^{\infty}B_{2n}\frac{1}{(2n)!}(e_*^{iw})^{2n}
\end{equation}
The uniqueness of inverses is ensured by Theorem\,\ref{nicefield}.

\medskip
On the other hand, replacing $z$ by $iw$ and using the formal power 
series of $(iw)^n_*$, we have also the identity 
\begin{equation}\label{Bern00}
\frac{1}{2}\Big(\sum_n\frac{(iw)_*^n}{(n+1)!}\Big)^{-1}
{+}\frac{1}{2}\Big(\sum_n\frac{(-iw)_*^n}{(n+1)!}\Big)^{-1}=
\sum_{n=0}^{\infty}
B_{2n}\frac{(iw)_*^{2n}}{(2n)!}
\end{equation}
where the r.h.s. is a formal power series of $(iw)_*^n$. 
This holds in every expression parameter $\tau$, and the r.h.s. may be 
regarded as a formal power series of
${:}(iw)_*^{2n}{:}_{\tau}$. 

As in \eqref{Euler22} it is very interesting to compare this with 
${iw}{*}(e_*^{iw}{-}1)^{-1}_{*+}{-}{iw}{*}(e_*^{-iw}{-}1)^{-1}_{*+}$ 
under the condition ${\rm{Re}}\,\tau>0$. However, 
this is written in the form 
${iw}{*}g_*(w)$ by using a $2\pi$-periodic function
$g_*(w)=g_*(w{+}2\pi)$, while the r.h.s. of \eqref{Bern00} has no such 
property.
Hence, we have to rewrite the l.h.s. of \eqref{Bern00} 
may be viewed as 
$\frac{iw}{2}{*}\Big((e_*^{iw}{-}1)^{-1}{-}(e_*^{-iw}{-1})^{-1}\Big)$. 

At this moment, we do not have an effective method to compare these,
but we will come back again to this interesting problem in the forthcoming paper.

\section{Fourier transform of tempered distributions}

We first recall the definition of rapidly 
decreasing functions of several variables.

\begin{definition}
  \label{rapdec}
A $C^{\infty}$-function, $f$ on $\Bbb R^n$, is called a 
rapidly decreasing function if 
$$
\sup_{x\in\Bbb R^n}|p(x)\partial^{\alpha}f(x)|<\infty
$$ 
for every polynomial $p(x)$ and for every multi-index $\alpha$. 
\end{definition}
We denote by $\Cal S(\Bbb R^n)$ the space of all rapidly decreasing 
functions. 
$\Cal S(\Bbb R^n)$ is a Fr{\'e}chet space under the family of 
seminorms 
$$
\|f\|_{k,m}=
\sum_{|\alpha|\leq m}{\text{sup}}(1+|x|^2)^k|\partial^{\alpha}f|
$$

First, express this space as the projective limit space 
of a family of  Hilbert spaces.
Define a family of inner products on 
$\Cal S(\Bbb R^n)$ as follows:

$$
\langle f,g\rangle_{k}=
\sum_{|\alpha|+|\beta|\leq k}
\int x^{2\beta}\partial^{\alpha}f\partial^{\alpha}g \dbar x 
$$
Make the topological completion of $\Cal S(\Bbb R^n)$ 
using the norm topology defined by 
the inner product, and denote it by $\Cal S^k(\Bbb R^n)$.
The Sobolev lemma gives that 
$\Cal S(\Bbb R^n)=\bigcap_k \Cal S^k(\Bbb R^n)$.

Fourier transform is defined as follows:
$$
\frak F(f)(\xi)=\int_{\Bbb R^n}f(x)e^{-ix\xi}\dbar x 
$$
where $\dbar x=\frac{1}{\sqrt{2\pi}}dx$. 
$\frak F(f)(\xi)$ is sometimes denoted by $\hat f(\xi)$.
Fourier transform is defined for $L^1$-functions at first, 
and extends in various ways. 
\begin{lem}
  \label{Planch}
If $f$, $f'$, $f''$ are continuous and summable, then  $\hat f(\xi)$ is 
summable, and it holds that $||f||_{L^2}=||\frak F(f)||_{L^2}$. 
\end{lem}

Fundamental properties of 
Fourier transform are as follows:
$$
\frak F(\partial^{\alpha}f)=(-i\xi)^{\alpha}\frak F(f), \qquad
\frak F((ix)^{\alpha}f)=\partial^{\alpha}\frak F(f).
$$
These are all proved by integration by parts.

Fourier transform exchanges differentiation and multiplication 
by generators. This observation suggests that 
the most convenient topology 
for the function space on which 
Fourier transform is defined is the topology where 
multiplication and differentiation by coordinate 
functions are treated with the same weight. 

It is well known that the 
Fourier transform  $\frak F$ gives a topological isomorphism of 
$\Cal S(\Bbb R_x^n)$ onto $\Cal S(\Bbb R_{\xi}^n)$. 
Furthermore, it gives a topological isomorphism of 
$\Cal S^k(\Bbb R_x^n)$ onto $\Cal S^k(\Bbb R_{\xi}^n)$ for every 
$k$. Setting 
$\Cal S(\Bbb R^n)=\bigcap_k \Cal S^k(\Bbb R^n)$, it gives  
an isomorphism of $\Cal S(\Bbb R^n)$. 

Thus, $\frak F$ gives a topological isomorphism of the dual space 
$\Cal S'(\Bbb R_{\xi}^n)$ onto $\Cal S'(\Bbb R_x^n)$. 

Let $\Cal S^{-k}(\Bbb R_x^n)$ be the dual space of 
$\Cal S^{k}(\Bbb R_x^n)$. 
The dual space of a Hilbert space is a Hilbert space 
by virtue of the Riesz theorem. 
Hence, $\Cal S^{-k}(\Bbb R_x^n)$ is a Hilbert space. We 
see easily that 
$$
\Cal S^{-\infty}(\Bbb R_x^n)=
\bigcup_{k}\Cal S^{-k}(\Bbb R_x^n)
$$ 
with the inductive limit topology.

\begin{lem}\label{FT-exp w}
If ${\rm{Re}}\tau >0$, then 
${:}\delta_*(z{+}w){:}_{\tau}=
\int_{\mathbb R}e^{itz}
e^{itw-\frac{\tau}{4}t^2}dt$ 
is an entire function of $z+w$.  
\end{lem}

Elements of  $\Cal S^{-\infty}(\Bbb R_x^n)$ 
are called {\bf tempered distributions}. 
If a tempered distribution $f$ is a function, that is, the value 
$f(x)$ is defined for every $x$, then 
$f$ is called a {\bf slowly increasing function}.

\begin{lem}\label{temparete}
For a polynomial $p(x)$ with real coefficients, 
$e^{ip(x)}$ is a slowly increasing function.
\end{lem}

\medskip

By Riesz's theorem  $\frak F$ extends to a linear isometry of  
 $\Cal S^{-k}(\Bbb R_x^n)$ onto $\Cal S^{-k}(\Bbb R_{\xi}^n)$. 
Hence, the Fourier transform $\frak F$ extends to 
a linear isomorphism of $\Cal S^{-\infty}(\Bbb R_x^n)$ onto  
$\Cal S^{-\infty}(\Bbb R_{\xi}^n)$.

\bigskip
Recall that if ${\rm{Re}}\,\tau>0$, then
${:}\delta_*(x{-}w){:}_{\tau}=
\frac{1}{\sqrt{\pi\tau}}e^{{-}\frac{1}{\tau}(x-w)^2}$ is rapidly decreasing.
Suppose $f(x)$ is $e^{|x|^{\alpha}}$-growth on ${\mathbb R}$ with 
$0{<}\alpha{<}2$.  
Then the integral $\int f(x){:}\delta_*(x{-}w){:}_{\tau}dx$ 
is well-defined to give an entire function w.r.t. $w$. 

The next theorem is the main tool to extend the class of 
$*$-functions via Fourier transform.
\begin{thm}
  \label{funddefm}
Suppose ${\rm{Re}}\,\tau{>}0$.
For every tempered distribution $f(x)$, 
the $\tau$-expression of  
$$
\int_{-\infty}^{\infty}f(x)\delta_*(x{-}w)dx
$$ 
is an entire function of $w$.  In particular we see 
$\delta_*(a{-}w)=\int_{-\infty}^{\infty}\delta(x{-}a)\delta_*(x{-}w)dx.$
\end{thm} 

\noindent
{\bf Proof}\,\,The $\tau$-expression is 
$\int_{-\infty}^{\infty}f(t)e^{-\frac{\tau}{4}t^2{+}itw}dt$. 
By restricting $w$ to an arbitrary compact subset of 
$\mathbb C$, $e^{-\frac{\tau}{4}t^2{+}itw}$ is rapidly decreasing 
w.r.t. $t$. 
Hence, the integral 
$\int_{-\infty}^{\infty}f(t)e^{-\tau t^2{+}itw}dt$ exists 
 for every tempered distribution $f(t)$. 
Since the complex differentiation $\partial_w$ does not suffer 
the convergence, we see that this is holomorphic 
on the whole plane. \hfill $\Box$

For every tempered distribution $f(x)$, 
we define a $*$-function $f_*(w)$ by 
\begin{equation}\label{startemparete}
f_*(w){=}\int_{-\infty}^{\infty}f(x)\delta_*(x{-}w)dx= 
\frac{1}{\sqrt{2\pi}} 
\int_{-\infty}^{\infty}\check f(t)e_*^{-itw}dt.
\end{equation}
where $\check f(t)$ is the inverse Fourier 
transform of $f(x)$. 
As $f(x)$ is a tempered distribution, one may write  
$$
\int f(x){:}\delta_*(x{-}w){:}_{\tau}dx 
= 
\frac{1}{2\pi}
\iint f(x)e^{itx}{:}e_*^{-itw}{:}_{\tau}dtdx
$$
under the existence of a rapidly decreasing function 
${:}e_*^{-itw}{:}_{\tau}$ in the integrand. 
By the definition of Fourier transform of tempered distribution, 
one may exchange the order of integrations. Hence, 
letting ${\check f}(t)$ be the inverse Fourier transform of 
$f(x)$, we have 
\begin{equation}\label{nicerep}
{:}\int_{\mathbb R} f(x)\delta_*(x{-}w)dx{:}_{\tau} 
= 
\frac{1}{\sqrt{2\pi}}
\int_{\mathbb R}{\check f}(t){:}e_*^{-itw}{:}_{\tau}dt
{=}{:}f_*(w){:}_{\tau}. 
\end{equation}

If another $*$-function is given by 
$g_*(w)=\int g(x)\delta_*(x{-}w)dx$, we define the 
product by 
\begin{equation}\label{productstar}
f_*(w){*}g_*(w)=
\int_{-\infty}^{\infty}f(w)g(w)\delta_*(x{-}w)dx
=\frac{1}{\sqrt{2\pi}}\int
\big(\frac{1}{\sqrt{2\pi}}\int {\check f}(t{-}\sigma)
{\check g}(\sigma)d\sigma\big)e_{*}^{-itw}dt,
\end{equation} 
if  $f(x)g(x)$ is defined as a tempered distribution or the convolution product 
$$
({\check f}\bullet{\check g})(t){=}
\frac{1}{\sqrt{2\pi}}\int {\check f}(t{-}\sigma){\check  g}(\sigma)d\sigma
$$ 
is defined as a tempered distribution. 
Hence, \eqref{productstar} may be viewed as an integral representation of
the intertwiner $I_0^{\tau}(f(x))=f_*(w)$.
If $f(x)$ is a slowly increasing function,  
applying \eqref{productstar} to the case $g_*(w)=\delta_*(a{-}w)$ gives 
\begin{equation}\label{funnystardelta}
f_*(w)*\delta_*(a{-}w)=
\int f(x)\delta(a{-}x)\delta_*(x{-}w)dx
{=}f(a)\delta_*(a{-}w).
\end{equation}

\subsection{Several applications}

Note that $\frac{1}{a-x}$, $a\not\in{\mathbb R}$, is a slowly
increasing function. If ${\rm{Im}\,a}<0$, then 
$\frac{1}{a-x}=\int_{-\infty}^{0}e^{ita}e^{-itx}dt$. Regarding this as
the Fourier transform of $\chi_{-}(t)e^{ita}$ supported on the
negative half-line, we have 
$\int\frac{1}{a-x}\delta_*(x{-}w)=\int_{-\infty}^{0}e^{ita}e_*^{-itw}dt=(a{-}w)_{*+}^{-1}$.
It is not hard to verify 
$$
\int\frac{1}{a-x}\delta_*(x{-}w)dx=
\left\{
\begin{matrix}
\medskip
(a{-}w)_{*+}^{-1}& {\rm{Im}\,a}<0\\
(a{-}w)_{*-}^{-1}& {\rm{Im}\,a}>0
\end{matrix}
\right.,\quad {\rm{Re}}\,\tau>0.
$$

Although the product 
$\delta_*(x{-}w){*}\delta_*(x{-}w)$ diverges, 
the next one is important 
\begin{equation}\label{sekidlt}
\delta_*(x{-}w){*}\delta_*(x'{-}w)=
\delta(x-x')\delta_*(x'{-}w) 
\end{equation} 
in the sense of distribution. This is proved directly as follows: 
$$
\begin{aligned}
\delta_*(x{-}w)&{*}\delta_*(x'{-}w)=
(\frac{1}{2\pi})^2\iint 
e_*^{it(x{-}w)}{*}e_*^{is(x{-}w)}dtds\\
=&
(\frac{1}{2\pi})^2\iint e^{itx{+}isx'}e_*^{-i(t{+}s)w}dtds=
(\frac{1}{2\pi})^2\iint e^{is(x'{-x)}}e_*^{i\sigma(x{-}w)}dsd\sigma
=\delta(x'{-}x)\delta_*(x{-}w).
\end{aligned}
$$
Note the trick that the computation is done without showing the
expression parameters. It is important to confirm this is true in the 
ordinary calculation, and we do not use the operator valued distributions. 

For ${\rm{Re}}\,\tau>0$, we define  
$$
Y_*(w)=\lim_{\e\downarrow 0}\int_{\e}^{\infty}\delta_*(x{-}w)dx, \quad 
Y_*(-w)=\lim_{\e\downarrow 0}\int_{-\infty}^{-\e}\delta_*(x{-}w)dx.
$$
It is clear that 
$$
\partial_w\lim_{\e\downarrow 0}\int_{\e}^{\infty}\delta_*(x{-}w)dx=
-\lim_{\e\downarrow 0}\int_{\e}^{\infty}\partial_x\delta_*(x{-}w)dx=
\delta_*({-}w)=\delta_*(w).
$$
By using \eqref{sekidlt} we have 
$$
Y_*(w){*}Y_*(w)=Y_*(w), \quad Y_*(w){*}Y_*(-w)=0,\quad 
Y_*(w){+}Y_*(-w)=\int_{\mathbb R}\delta_*(x{-}w)dx=1.
$$
We define 
$$
{\rm{sgn}}_*(w)=Y_*(w){-}Y_*(-w).
$$
It is easy to see that  
${\rm{sgn}}_*(w){*}{\rm{sgn}}_*(w)=Y_*(w){+}Y_*(-w)=1, \quad {\rm{sgn}}_*(w){+}{\rm{sgn}}_*(-w)=0. $

\bigskip
Since $\delta_*(z{-}w)$ is holomorphic in $z$, Cauchy 
integral theorem gives that every contour integral vanishes, but we
see easily  for every simple closed curve $C$
$$
\frac{1}{2\pi i}\int_C \frac{1}{z}\delta_*(z{-}w)dz=\delta_*(w),\quad
{\rm{Re}}\tau >0.
$$  
\medskip
Note that ${\rm{v.p.}}\frac{1}{x}$, $\rm{Pf}.x^{-m}, m\in \mathbb N$ are tempered
distribution which are not functions, but their Fourier transform may
be viewed as slowly increasing functions. Hence we see 
$$
\begin{aligned}
{\rm{v.p.}}\int_{\mathbb R}\frac{1}{x}\delta_*(x{-}w)dx&= 
\frac{-i}{\sqrt{2\pi}}
\int_{\mathbb R}\!\sqrt{\frac{\pi}{2}}\,{\rm{sgn}}(t)e_*^{-itw}dt
=\frac{1}{2}(w_{*+}^{-1}{+}w_{*-}^{-1})\\ 
{\rm{Pf}.}\int_{\mathbb R}x^{-m}\delta_*(x{-}w)dx&=
\frac{-i}{2}
\int_{\mathbb R}\frac{1}{(m{-}1)!}(-it)^{m{-}1}{\rm{sgn}}(t)e_*^{-itw}dt
=(-1)^{m-1}\frac{1}{2}(w_{*+}^{-m}{+}w_{*-}^{-m}).
\end{aligned}
$$

\subsubsection{Periodic distributions}
A tempered distribution $f(x)$ is called a $2\pi$-periodic 
tempered distribution, if $f(x)$ satisfies $f(x{+}2\pi)=f(x)$.  
For every distribution $f(x)$ of compact support, the infinite sum 
$\sum_nf(x{+}2\pi n)$ is a $2\pi$-periodic tempered distribution. 
This procedure will be called the $2\pi$-{\it periodization}.

\bigskip
As $f(t)=\sum_ne_*^{i(n{+}t)(a{+}w)}$ is a periodic function of period
$1$, Fourier expansion formula gives 
$f(t)=\sum_m(\int_0^{1} f(s)e^{-2\pi ims}ds)e^{2\pi imt}$ and 
$$ 
f(0)=\sum_m(\int_0^1\sum_ne_*^{i(n{+}s)(a{+}w)}e^{-2\pi ims}ds)=
\int_0^1\sum_m\sum_ne_*^{i(n{+}s)(a{+}w)}e^{-2\pi ims}ds
$$
As $e_*^{i(n{+}s)(a{+}w)}e^{-i2\pi ms}=e_*^{i(n{+}s)(a{+}w)}e^{-i(2\pi m(n{+}s)}$,
$$
\int_0^1\sum_m\sum_ne_*^{i(n{+}s)(a{+}w)}e^{-2\pi ims}ds{=}
\sum_m\int_{\mathbb R}e_*^{is(a{+}w{+}2\pi m)}ds{=}
\sum_m\delta_{*}(a{+}w{+}2\pi m).
$$
It follows the fundamental relation between 
$2\pi$-periodic tempered distributions and 
Fourier series 
\begin{equation}\label{periodic}
\sum_n\delta_*(a{+}2\pi n{+}w)=\sum_ne_*^{in(a{+}w)}.
\end{equation}

A continuous function $f(x)$ on $[-\pi,\pi]$ extends to a (not continuous) 
$2\pi$-periodic function ${\tilde f}_{\pi}(x)$ to give a 
$2\pi$-periodic tempered distribution, where 
$$
{\tilde f}_{\pi}(x)=\frac{1}{2\pi}\sum_n(\int_{-\pi}^{\pi}f(s)e^{-ins}ds)e^{inx}
=\sum_na_ne^{inx}
$$
Hence, 
\begin{equation}\label{perperper}
{\tilde f}_{\pi *}(w)=\int_{\mathbb R}{\tilde f}_{\pi}(x)\delta_*(x{-}w)dx=\sum a_ne_*^{inw}.
\end{equation} 
In particular by using $\frac{1}{in}(x^me^{-inx})'=\frac{1}{in}mx^{m{-}1}e^{-inx}{-}x^me^{-inx}$,
we have 
$$
{\tilde x}_{\pi *}(w)=
\int_{\mathbb R}{\tilde x}_{\pi}(x)\delta_*(x{-}w)dx
=\frac{1}{2\pi}\sum_{n=-\infty}^{\infty}\frac{i}{n}(-1)^{n{-}1}e_*^{inw}.
$$
$$
{\tilde x}^2_{\pi *}(w)=\int_{\mathbb R}{\tilde x}^2_{\pi}(x)\delta_*(x{-}w)dx
=\frac{1}{2\pi}\sum_{n=-\infty}^{\infty}\frac{2}{n^2}(-1)^{n{-}1}e_*^{inw}
$$

\begin{lem}\label{solution}
If a periodic distribution $f_*(w)$ satisfies 
${\tilde x}_{\pi *}(w){*}f_*(w)=0$, then 
$$
f_*(w)=c\sum_n\delta_*(w{+}2\pi n)=c\sum_ne_*^{inw}.
$$
\end{lem}

\noindent
{\bf Proof}\,\, 
Note that  ${\tilde x}_{\pi *}(w){*}f_*(w)=
\int {\tilde x}_{\pi}(x)f(x)\delta_*(x{-}w)dx$.  
Hence, ${\tilde x}_{\pi}(x)f(x)=0$ where $f(x)$ is a periodic distribution.   
As ${\tilde x}_{\pi}(x)$ is a $2\pi$-periodization of $x$, this means 
$xf(x)=0$ on a neighborhood of $0$. This gives 
$f(x)=c\delta(x)$ in a neighborhood of $0$. The periodicity of $f(x)$
 gives $f(x)=\sum_k\delta(x{+}2\pi k)$ and hence, 
$f_*(w)=\sum_k\delta_*(w{+}2\pi k)$. \hfill $\Box$.

\subsection{Double-valued parallel sections}\label{resoventcalculus}

Proposition\,\ref{unique} shows that the equation 
${:}e_*^{2iw}{:}_{\tau}{*_{\tau}}f(w)=1$ has the unique solution $e_*^{-2iw}$.
On the other hand, two inverses $(a{+}w)_{*\pm}^{-1}$ are defined for
every $a\in{\mathbb C}$ by the integral 
$$
{:}(a{+}w)_{*+}^{-1}{:}_{\tau}= 
i\int_{-\infty}^0{:}e_*^{it(a+w)}{:}_{\tau}dt,\quad 
{:}(a{+}w)_{*-}^{-1}{:}_{\tau}= 
{-}i\int^{\infty}_0{:}e_*^{it(a+w)}{:}_{\tau}dt, 
\quad {\rm{Re}}\,\tau{>}0.
$$
So far, the expression parameter $\tau$ is mainly restricted to the
domain ${\rm{Re}}\,\tau>0$. 
However, for every $\tau{\not=}0$ 
there is $\theta$ such that $e^{2i\theta}\tau{>}0$. 
Keeping these in mind, we compute the $\tau$-expression of 
$$
e^{i\theta}\int^0_{-\infty}e_*^{ite^{i\theta}(a{+}w)}dt, 
\quad 
-e^{i\theta}\int_0^{\infty}e_*^{ite^{i\theta}(a{+}w)}dt. 
$$
For the first one, we see 
\begin{equation}
  \label{eq:intlm}
{:}\,e^{i\theta}\!\int^0_{-\infty}
e_*^{ite^{i\theta}(a{+}w)}dt{:}_\tau{=}
e^{i\theta}\!\int^0_{-\infty}\!
e^{ite^{i\theta}(a{+}w){-}\frac{1}{4}t^2e^{2i\theta}\tau}dt  
\end{equation}
converges absolutely to give the inverse $a{+}w$. 
Moreover, we see by integration by parts 
$$
\begin{aligned}
\frac{d}{d\theta}{:}\,e^{i\theta}\!\int^0_{-\infty}
e_*^{ite^{i\theta}(a{+}w)}dt{:}_\tau&=
{:}\,ie^{i\theta}\!\int^0_{-\infty}
e_*^{ite^{i\theta}(a{+}w)}dt{:}_\tau
{+}
{:}\,e^{i\theta}\!\int^0_{-\infty}
\frac{d}{d\theta}e_*^{ite^{i\theta}(a{+}w)}dt{:}_\tau\\
&=
{:}\,ie^{i\theta}\!\int^0_{-\infty}
e_*^{ite^{i\theta}(a{+}w)}dt{:}_\tau
{-}
{:}\,ie^{i\theta}\!\int^0_{-\infty}
e_*^{ite^{i\theta}(a{+}w)}dt{:}_\tau=0
\end{aligned}
$$ 
 whenever the property ${\rm{Re}}\,e^{2i\theta}\tau{>}0$ is retained. 
This is proved also by Cauchy's integration theorem via the 
rotation of the path of integration.

If the expression parameter is restricted 
in a simply connected domain $D$ of ${\mathbb C}{\setminus}\{0\}$, 
one can treat ${:}(a{+}w)_{*\pm}^{-1}{:}_{\tau}$ as a single valued 
parallel section on $D$. In such a restricted domain, products of inverse 
elements are given by the resolvent identities, which gives an 
associative algebra similar to the half-series algebra.

This observation suggests that the natural boundary of the 
expression parameter $\tau$ of the discrete summation 
such as $\theta_3(w,*)$ causes an essential 
difference between $\delta_*(w)$ and $\theta_3(w,*)$.

\medskip
The following proposition shows that 
${:}(a{+}w)_{*+}^{-1}{:}_{\tau}$ and    
${:}(a{+}w)_{*-}^{-1}{:}_{\tau}$  are connected 
by a parallel translation along a closed curve 
of expression parameters. But since  
parallel translations in this case are always considered 
with $\theta$ such that $|2\theta{+}\sigma|<\frac{\pi}{2}$, 
we refer such translation as 
{\bf joint parallel displacements/translations}.  
\begin{prop}\label{strnginv}
Suppose $|2\theta{+}\sigma|{<}\frac{\pi}{2}$. Then, the 
$|\tau|e^{i\sigma}$-expression of 
$e^{i\theta}\!\int^0_{-\infty}
e_*^{ite^{i\theta}(a{+}w)}dt$ is independent of $\theta$. 
That is, in the domain $(\theta,\sigma)$, the element 
given by the integral  
$$
{:}e^{i\theta}(e^{i\theta}(a{+}w))_{*+}^{-1}
{:}_{|\tau|e^{i\sigma}}
={:}\,e^{i\theta}\!\int^0_{-\infty}
e_*^{ite^{i\theta}(a{+}w)}dt{:}_{|\tau|e^{i\sigma}} 
$$
depends only on the expression parameter 
$\tau=|\tau|e^{i\sigma}$. Hence, it may be 
viewed as a single element. 
However, moving $\theta$ from $0$ to $-\frac{\pi}{2}$ 
in this domain, we have 
$$
{:}\,i\!\int^0_{-\infty}
e_*^{it(a{+}w)}dt{:}_{\tau}=
{:}\,\!\int^0_{-\infty}
e_*^{t(a{+}w)}dt{:}_{-\tau}.
$$ 
Moving $\theta$ from $0$ to $\pi$, we see 
$$
{:}\,\!\int^0_{-\infty}
e_*^{it(a{+}w)}dt{:}_{\tau}=
{:}\,{-}\!\int^0_{-\infty}
e_*^{-it(a{+}w)}dt{:}_{\tau}=
-{:}\,\!\int_0^{\infty}
e_*^{it(a{+}w)}dt{:}_{\tau}.
$$  
In particular, the complex rotation $e^{i\theta}$ from $\theta=0$ to
$\pi$ of the path of integration together 
with the expression parameter, exchanges 
$(a{+}w)_{*+}^{-1}$ and $(a{+}w)_{*-}^{-1}$. 
\end{prop}

\noindent
{\bf Remark}\,\, The last equality does not imply
$(a{+}w)_{*+}^{-1}=(a{+}w)_{*-}^{-1}$. It expresses only the 
result of parallel translation along a closed path.  
For  ${\rm{Re}}\,\tau >0$, these may be written as 
$$
(a{+}w)_{*+}^{-1}\,
{=}i\int_{-\infty}^0e_*^{it(a+w)}dt,\quad   
(a{+}w)_{*-}^{-1}\,
{=}{-}i\!\int_{0}^{\infty}e_*^{it(a+w)}dt,
$$
without showing the expression parameters explicitly. 

\bigskip
\mbox{\parbox{.4\linewidth}{
\setlength{\unitlength}{.25mm}
\begin{picture}(200,150)
\thinlines
\put(0,60){\line(1,0){200}}
\put(100,0){\line(0,1){130}}
\put(100,50){\line(2,-1){100}}
\put(100,50){\line(-2,1){100}}
\put(100,70){\line(2,-1){100}}
\put(100,70){\line(-2,1){100}}
\put(100,130){$\theta$}
\put(190,60){$\sigma$}
\put(105,58){$\bullet$}
\put(105,48){$\bullet$}
\thicklines
\put(100,60){\line(1,0){11}}
\put(110,60){\line(0,-1){10}}
\put(110,50){\line(1,0){18}}
\put(128,50){\line(0,-1){11}}
\put(128,40){\line(1,0){18}}
\put(146,40){\line(0,-1){11}}
\put(146,30){\line(1,0){18}}
\put(0,5){\footnotesize{Fig.1}}
\end{picture}
}}
\hfill{\parbox{.55\linewidth}{
Since 
${:}\,e^{i\theta}\!\int^0_{-\infty}
e_*^{ite^{i\theta}(a{+}w)}dt{:}_{|\tau|e^{i\sigma}}$ is 
viewed as various expressions of 
a single element defined on the domain 
$|2\theta{+}\sigma|{<}\frac{\pi}{2}$, it may be 
written by 
$(a+w)_*^{-1}$. However, this is a double-valued single parallel section.   
We do not use the notation
$(a+w)_*^{-1}$ as this may cause some confusion. 
In our expression, 
$(a{+}w)_{*\pm}^{-1}$ are viewed as parallel 
sections defined on the space of expression parameters 
sitting in the domain $|2\theta{+}\sigma|{<}\frac{\pi}{2}$. 
Proposition\ref{strnginv} shows how 
these expressions are related to each other.}  

\medskip
As 
$\delta_*(a{+}w)
=(a+w)^{-1}_{*+}-(a+w)^{-1}_{*-}$, 
Proposition\,\ref{strnginv} must give that 
$e^{i\theta}\delta_*(e^{i\theta}(a+w))$ depends only on the 
expression parameter on the domain 
$|2\theta{+}\sigma|{<}\frac{\pi}{2}$. Indeed, we see 
\begin{equation}\label{deltaindep}
{:}e^{i\theta}\delta_*(e^{i\theta}(a+w)){:}_{|\tau|e^{i\sigma}}=
\frac{1}{\sqrt{\pi|\tau|e^{i\sigma}}}
e^{-\frac{1}{|\tau|e^{i\sigma}}(a+w)^2}
\end{equation}
by the Fourier transform and this is rapidly decreasing. Setting
$\sigma=-2\theta$ and moving $\theta$ from $0$ to $\pi$, we see that 
$$
{-}{:}\delta_*(-(a+w)){:}_{\tau}=-\delta_*(a+w){:}_{\tau}=\delta_*(a+w){:}_{\tau}.
$$
\begin{prop}\label{multi-valued}
The mapping $\delta_*:{\mathbb C}{\setminus}\{0\}
\rightarrow H{\!o}l(\mathbb C)$
defined by $\tau{\to}{:}\delta_{*}{:}_{\tau}$ is  a double-valued 
parallel section with $\pm$-ambiguity.
\end{prop}

\bigskip
In contrast with Proposition\,\ref{unique}, we easily see the following:
\begin{prop}\label{delta22}  
For every $\tau\not=0$, the real analytic solution of the differential equation
$$
(a{+}w){*_{\tau}}f(w)=0
$$ 
is $f(w)=Ce^{-\frac{1}{\tau}(a+w)^2}$. That
is, the space of solutions form the trivial bundle 
${\mathbb C}_*{\times}{\mathbb C}e^{-\frac{1}{\tau}(a+w)^2}$, while
$C{:}\delta_*(a{+}w){:}_{\tau}=
  \frac{1}{\sqrt{\pi\tau}}e^{-\frac{1}{\tau}w^2}$ forms a M{\"o}bius bundle over 
${\mathbb C}_*$. 
\end{prop}

\noindent
{\bf Note}\,\,
$f(\theta,\sigma)=
{:}e^{i\theta}\delta_*(e^{i\theta}(a+w)){:}_{|\tau|e^{i\sigma}}$ is a
smooth function defined on the strip in Fig.1  such that 
$\partial_{\theta}f(\theta,\sigma)=0$. But it may loose the important 
information to write this $f(\theta,\sigma)=f(\sigma)$.
This implies that $\delta_*(a{+}w)$ must be treated as a 
parallel section of the M{\"o}bius bundle over ${\mathbb C}_*$, while 
the solution of $(a{+}w){*_{\tau}}f(w)=0$ can be treated as trivial
parallel sections of the product bundle.  
This is not a minor difference, but 
gives a big philosophical difference. 
It is natural to expect there is a ``more'' twisted bundles. 
In the last part we give a candidate for this by considering 
$*$-exponential function $e_*^{sw_*^{\ell}}$ for $\ell>3$.

\bigskip

Since $\int^{0}_{-\infty}e_*^{it(a{+}w)}dt$ and 
$-\int^{\infty}_{0}e_*^{it(a{+}w)}dt$ are connected by a parallel transform along 
a closed curve, these may be viewed as the same element. Though  
Proposition\,\ref{strnginv} insists these can never be 
distinguished globally, these two can be distinguished locally. 

\begin{center}
It is remarkable that there is a single object with 
many indistinguishable individuals. 
\end{center}

\bigskip
In spite of this, it is easy to see by change of variables that for every real 
$\lambda\not=0$, 
\begin{equation}\label{scale}
(\lambda(a+w))^{-1}_{*\pm}=\lambda^{-1}(a+w)^{-1}_{*\pm}, \quad 
\delta_*(\lambda(a+w))=\lambda^{-1}\delta_*(a+w), \quad 
\lambda\in {\mathbb R}\setminus\{0\}.
\end{equation}
For instance, replacing $t\lambda\to t'$ gives
$$
i\int_{-\infty}^0e_*^{it\lambda(a+w)}dt=i\lambda^{-1}\int_{-\infty}^0e_*^{it'(a+w)}dt'.
$$ 
On the other hand, we see that 
\begin{equation}\label{doublepi}
\int_{\mathbb R}\delta_*(x{+}w)dx=\frac{1}{\sqrt{\pi\tau}}\int_{\mathbb R}
e^{-\frac{1}{\tau}(x{+}w)^2}dx=
\frac{1}{\sqrt{\pi\tau}}
\int_{\mathbb R}e^{-\frac{1}{\tau}(x{+}w)^2}dw
=1, \quad {\rm{Re}}\,\tau >0,
\end{equation}
is independent of the expression parameter $\tau$.

\subsection{Complex rotations of $*$-functions}

Recall that $\delta_*(a+w)$ is double-valued. In precise,
${:}e^{i\theta}\delta_*(e^{i\theta}(a+w))
{:}_{|\tau|e^{i\sigma}}$ is a double-valued parallel section under the
joint parallel translation. Indeed, this is given by 
\begin{equation*}
{:}e^{i\theta}\delta_*(e^{i\theta}(a+w))
{:}_{|\tau|e^{i\sigma}}=
\frac{1}{\sqrt{\pi|\tau|e^{i\sigma}}}
e^{-\frac{1}{|\tau|e^{i\sigma}}(a+w)^2}
\end{equation*}
by the Fourier transform and it is rapidly decreasing on 
the domain $|2\theta{+}\sigma|<\frac{1}{2}\pi$. Note that the value dose
not depend on $\theta$. 
Hence, for every tempered distribution $f(x)$ the integral 
$$
{:}f_*(w){:}_{\tau e^{i\sigma}}
=\int_{\mathbb R}f(x){:}e^{i\theta}\delta_*(e^{i\theta}(x+w)){:}_{\tau e^{i\sigma}}dx
$$
is defined as a double-valued parallel section with $\pm$-ambiguity on
the domain $|2\theta{+}\sigma|<\frac{1}{2}\pi$.

\bigskip
An interesting phenomenon appears in the integral 
\begin{equation}\label{ellroots}
\int_{\mathbb R}e^{isx^{\ell}}\delta_*(x{-}w)dx, \quad 
s{\in}{\mathbb R}, 
\end{equation}
where $\ell$ is an integer $\ell\geq 3$. As $e^{isx^{\ell}}$ is slowly
increasing function of $x$, the integral converges and defined an
element which looks an element written as $e_*^{isw_*^{\ell}}$. 
We now consider the rotation $e^{i\theta}x$ of the path of integral 
from $0$ to $2\pi/\ell$ by keeping ${\rm{Re}}\,e^{2i\theta}\tau>0$. 
Then the integral 
$$
\int_{\mathbb R}e^{is(e^{i\theta}x)^{\ell}}\delta_*(e^{i\theta}x{-}w)d(e^{i\theta}x)
$$
changes into 
$$
e^{i\frac{2\pi}{\ell}}\int_{\mathbb R}e^{isx^{\ell}}\delta_*(x{-}w)dx.
$$
Thus we see what is defined by \eqref{ellroots} is not 
$e_*^{isw_*^{\ell}}$ but something like its $\ell$-th root. 
Recalling that $e^{i\theta}\delta_*(e^{i\theta}x{-}w)$ is 
independent of $\theta$, we see that it is impossible to distinguish 
$$
e^{i\frac{2\pi k}{\ell}}
\int_{\mathbb R}e^{isx^{\ell}}\delta_*(x{-}w)dx,\quad k=0,1,2,\cdots,\ell{-}1.
$$

The multi-valued nature makes it difficult to define 
$$
{:}f_*(w){:}_{\tau e^{i\sigma}}{+}{:}g_*(w){:}_{\tau e^{i\sigma}}
$$
if the expression parameter can move independently. Thus it is better
to define 
$$
{:}f_*(w){+}g_*(w){:}_{\tau e^{i\sigma}}
=\int_{\mathbb R}(f(x){+}g(x)){:}e^{i\theta}\delta_*(e^{i\theta}(x+w)){:}_{\tau e^{i\sigma}}dx.
$$

\subsubsection{Double-valued nature disappears under 
integration}

If ${\rm{Re}}\,\tau >0$, we easily see that    
$\int_{\mathbb R}{:}\delta_{*}(a{-}w){:}_{\tau}dw=1$,  
independent of the expression parameter $\tau$. 
Hence, \eqref{funnystardelta} shows also 
$$
\int_{\mathbb R}{:}f_*(w)*\delta_*(a{-}w){:}_{\tau}dw= f(a), 
$$
independent of the expression parameter $\tau$.
Similarly, we have 
$$
\frac{1}{e^{i\theta}\sqrt{\pi\tau}}\int_{\mathbb R}
e^{-e^{-2i\theta}\frac{1}{\tau}
(e^{i\theta}x+w)^2}d(e^{i\theta}x) 
=\frac{1}{\sqrt{\pi\tau}}\int_{\mathbb R}
e^{-\frac{1}{\tau}(x+e^{-i\theta}w)^2}dx =1.
$$ 
Keeping this in mind, we define 
\begin{equation}
\int_{\mathbb R}{:}f_*(w){*}e^{i\theta}\delta_*(e^{i\theta}(x+w)){:}_{\tau e^{i\sigma}}dw= f(a)
\end{equation}
and call this the {\bf formal $*$-integration}.

\subsubsection{Another $*$-inverse of $\sin_*\pi w$.}
A rotations of path of integration gives sometimes a powerful tool of 
calculation. Here we give an example, though this is not directly
relevant to our purpose.

Recall that the classical formula  
\begin{equation}\label{eq:sincos100}
\sin\pi z
\int_{-\infty}^{\infty}
\frac{e^{tz}}
{1{+}e^{t}}dt=\pi=\varGamma(z)\varGamma(1{-}z)
\end{equation} 
obtained easily by Cauchy's integral formula. The proof may be applied
to our case to obtain an inverse of $\sin_*w$: Consider first the integral 
$(e_*^{2\pi i(z{+}w)}{-}1)*\!
\int_{-\infty}^{\infty}\frac{e_*^{t(z{+}w)}}{1{+}e^t}dt$ for the case 
${\rm{Re}}\,\tau <0$.  
Including $(e_*^{2\pi i(z{+}w)}{-}1)$ among the integrand,  
we have integrals along $2\pi i{+}{\mathbb R}$ 
to positive direction and along ${\mathbb R}$ to the 
negative direction. Hence, by Cauchy integration 
formula, we have the negative residue: 
$$
(e_*^{2\pi i(z{+}w)}{-}1)*\!
\int_{-\infty}^{\infty}\frac{e_*^{t(z{+}w)}}{1{+}e^t}dt
{=}2\pi i e_*^{\pi i(z{+}w)},\quad ({\rm{Re}}\,\tau<0). 
$$
Multiplying $e_*^{-\pi i(z{+}w)}$ to both sides by using 
the associativity \eqref{eq:prodexp22}, we have 
$$
2i\sin_*\pi(z{+}w)*\!
\int_{-\infty}^{\infty}\frac{e_*^{t(z{+}w)}}{1{+}e^t}dt{=}
2\pi i
$$
This gives another $*$-inverse  
of $\sin_*\pi(z{+}w)$ different from 
$(\sin_*\pi(z{+}w))^{-1}_{*\pm}$ given by the theta function, since the latter is defined for 
${\rm{Re}}\,\tau >0$.  

\medskip
To investigate the relation between these inverses, we have to rotate the expression
parameter. Note that Cauchy's integration theorem gives for every $-\pi<a<\pi$ 
$$
f_*(w)=\int_{-\infty}^{\infty}\frac{e_*^{tw}}{1{+}e^t}dt
=\int_{-\infty}^{\infty}\frac{e_*^{(t{+}ia)w}}{1{+}e^{t{+}ia}}dt
=\int_{-\infty}^{\infty}
\frac{e_*^{tw}}{e^{-ia}{+}e^{t}}dt{*}e_*^{ia(w{-}1)}.
$$
The r.h.s. does not depend on $a$. 

Fix $\tau$ so that ${\rm{Re}}\,\tau<0$. 
Consider $e^{i\sigma}\tau$-expression instead of 
$\tau$-expression together with the $e^{i\theta}$-
rotation of the path of integral   
by keeping ${\rm{Re}}\,e^{i(2\theta+\sigma)}\tau<0$, 
and set
$$
{:}g_{*\theta}(w){:}_{e^{i\sigma}\tau}=
{:}\int_{-\infty}^{\infty}\frac{e_*^{e^{i\theta}tw}}
{e^{-ia}{+}e^{e^{i\theta}t}}d(e^{i\theta}t){:}_{e^{i\sigma}\tau}, 
\quad 
{\rm{Re}}\,e^{2i\theta}e^{i\sigma}\tau<0.
$$
Since there is no singular point in a small sector, 
the rotation of the path 
changes nothing. This is also confirmed by checking 
$$
\frac{d}{d\theta}
\int_{-\infty}^{\infty}\frac{e_*^{e^{i\theta}tw}}
{e^{-ia}{+}e^{e^{i\theta}t}}d(e^{i\theta}t){=}0
$$
via integration by parts. We set $g_{*\theta}(w){*}e_*^{ia(w{-}1)}=f_{*\theta}(w)$.
Since  $f(w)=f_{*\theta}(w)=g_{*\theta}(w){*}e_*^{ia(w{-}1)}$, 
we also have the identities
$$
f_{*\theta}(w)=f_{*\theta}(1{-}w),\quad 
f_{*\theta}(w){+}f_{*\theta}(w{+}1)=
\int_{-\infty}^{\infty}e_*^{e^{i\theta}tw}d(e^{i\theta}t)=e^{i\theta}\delta_*(-e^{i\theta}w).
$$
Hence, we have 
\begin{equation}\label{eq:sincos00}
\frac{1}{\pi}\sin_*\pi w*
\int_{-\infty}^{\infty}\frac{e_*^{e^{i\theta}tw}}
{1{+}e^{e^{i\theta}t}}d(e^{i\theta}t)=1.
\end{equation}

Taking the path of integration as in $\mu$ and $\nu$ in 
the l.h.s. figure below. Then, it is easy to see
$$
2f_*(w)=
\int_{\mu}
\frac{e_*^{(t{+}ia)w}}{1{+}e^{t{+}ia}}dt
{+}
\int_{\nu}
\frac{e_*^{(t{-}ia)w}}{1{+}e^{t{-}ia}}dt.
$$

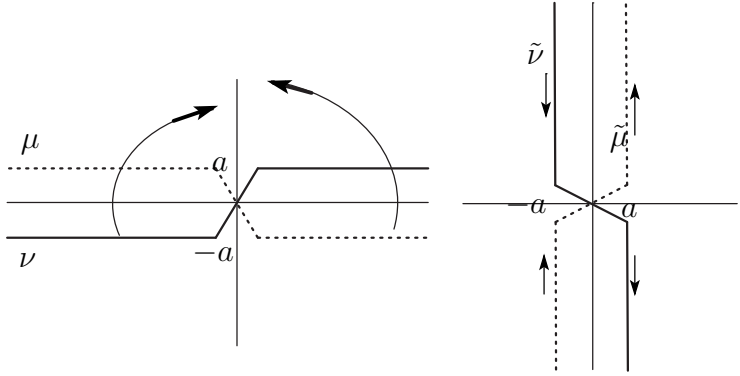
\unitlength 0.1in
\begin{picture}( 37.8000, 19.2000)(  1.8000,-19.7000)
%
\special{pn 8}%
\special{pa 180 1092}%
\special{pa 2360 1092}%
\special{fp}%
\special{pa 1370 450}%
\special{pa 1370 1840}%
\special{fp}%
%
\special{pn 13}%
\special{pa 190 914}%
\special{pa 1260 914}%
\special{dt 0.045}%
\special{pa 1260 914}%
\special{pa 1478 1276}%
\special{dt 0.045}%
\special{pa 1478 1276}%
\special{pa 2350 1276}%
\special{dt 0.045}%
%
\special{pn 13}%
\special{pa 180 1276}%
\special{pa 1260 1276}%
\special{fp}%
\special{pa 1260 1276}%
\special{pa 1478 914}%
\special{fp}%
\special{pa 1478 914}%
\special{pa 2360 914}%
\special{fp}%
%
\special{pn 13}%
\special{pa 3384 58}%
\special{pa 3390 1000}%
\special{dt 0.045}%
\special{pa 3390 1000}%
\special{pa 3020 1194}%
\special{dt 0.045}%
\special{pa 3020 1194}%
\special{pa 3022 1964}%
\special{dt 0.045}%
%
\special{pn 13}%
\special{pa 3014 50}%
\special{pa 3018 1002}%
\special{fp}%
\special{pa 3018 1002}%
\special{pa 3390 1194}%
\special{fp}%
\special{pa 3390 1194}%
\special{pa 3394 1970}%
\special{fp}%
%
\special{pn 8}%
\special{pa 2540 1098}%
\special{pa 3960 1098}%
\special{fp}%
\special{pa 3212 1964}%
\special{pa 3212 50}%
\special{fp}%
\special{pa 3212 50}%
\special{pa 3212 50}%
\special{fp}%
%
\special{pn 8}%
\special{ar 1370 1092 832 650  4.9284453 6.2831853}%
\special{ar 1370 1092 832 650  0.0000000 0.2155411}%
%
\special{pn 8}%
\special{ar 1370 1100 644 502  2.8104084 4.5204397}%
%
\special{pn 20}%
\special{pa 1756 520}%
\special{pa 1558 466}%
\special{fp}%
\special{sh 1}%
\special{pa 1558 466}%
\special{pa 1616 502}%
\special{pa 1608 480}%
\special{pa 1628 464}%
\special{pa 1558 466}%
\special{fp}%
\special{pa 1558 466}%
\special{pa 1558 466}%
\special{fp}%
%
\special{pn 20}%
\special{pa 1042 666}%
\special{pa 1240 598}%
\special{fp}%
\special{sh 1}%
\special{pa 1240 598}%
\special{pa 1170 600}%
\special{pa 1190 616}%
\special{pa 1184 638}%
\special{pa 1240 598}%
\special{fp}%
\put(12.4000,-9.3000){\makebox(0,0)[lb]{$a$}}%
\put(11.4000,-14.1000){\makebox(0,0)[lb]{$-a$}}%
\put(33.6000,-11.7000){\makebox(0,0)[lb]{$a$}}%
\put(27.6000,-11.6000){\makebox(0,0)[lb]{$-a$}}%
\put(2.5000,-8.4000){\makebox(0,0)[lb]{$\mu$}}%
\put(2.4000,-14.3000){\makebox(0,0)[lb]{$\nu$}}%
%
\special{pn 8}%
\special{pa 2970 420}%
\special{pa 2970 420}%
\special{fp}%
\special{pa 2970 420}%
\special{pa 2970 640}%
\special{fp}%
\special{sh 1}%
\special{pa 2970 640}%
\special{pa 2990 574}%
\special{pa 2970 588}%
\special{pa 2950 574}%
\special{pa 2970 640}%
\special{fp}%
%
\special{pn 8}%
\special{pa 3430 740}%
\special{pa 3430 480}%
\special{fp}%
\special{sh 1}%
\special{pa 3430 480}%
\special{pa 3410 548}%
\special{pa 3430 534}%
\special{pa 3450 548}%
\special{pa 3430 480}%
\special{fp}%
%
\special{pn 8}%
\special{pa 3430 1370}%
\special{pa 3430 1570}%
\special{fp}%
\special{sh 1}%
\special{pa 3430 1570}%
\special{pa 3450 1504}%
\special{pa 3430 1518}%
\special{pa 3410 1504}%
\special{pa 3430 1570}%
\special{fp}%
%
\special{pn 8}%
\special{pa 2960 1570}%
\special{pa 2960 1370}%
\special{fp}%
\special{sh 1}%
\special{pa 2960 1370}%
\special{pa 2940 1438}%
\special{pa 2960 1424}%
\special{pa 2980 1438}%
\special{pa 2960 1370}%
\special{fp}%
\put(28.7000,-3.7000){\makebox(0,0)[lb]{$\tilde{\nu}$}}%
\put(33.0000,-8.3000){\makebox(0,0)[lb]{$\tilde{\mu}$}}%
\end{picture}%
\hfill
\parbox[b]{.23\linewidth}
{Set $\theta=-\frac{1}{2}\sigma$ and move $\sigma$ 
from $0$ to $2\pi$ in the path $\mu$, and $0$ to $-2\pi$ in 
the path $\nu$. 
At $\sigma=\pi$, $-\pi$, the expression parameter is $-\tau$
and the path of integrals are changed as $\tilde{\mu}$ and 
$\tilde{\nu}$ avoiding singular points. Thus, $2f_*(w)$ is 
changed into :
}

$$
{:}f_{*{-}\frac{\pi}{2}}(w){:}_{-\tau}+
{:}f_{*\frac{\pi}{2}}(w){:}_{-\tau}
=
{:}\int_{\tilde\mu}\frac{e_*^{-itw}}{1{+}e^{-it}}d(-it){:}_{-\tau}
+
{:}\int_{\tilde\nu}\frac{e_*^{itw}}
{1{+}e^{it}}d(it){:}_{-\tau}. 
$$
Note that the alternating periodicity $w\to w{+}1$ appears in the r.h.s. because
the paths of integrations are reversed in the first and second term of
the r.h.s.

Switch the path of integration into upper and lower circuits.  
We now count the residues in the upper circuits and lower circuits separately.  Then 
$$
{:}f_{*{-}\frac{\pi}{2}}(w){:}_{-\tau}+
{:}f_{*\frac{\pi}{2}}(w){:}_{-\tau}
=2\pi\sum_{n=0}^{\infty}
{:}e_*^{\pi i(2n+1)w}{:}_{-\tau} 
-
2\pi\sum_{n=0}^{\infty}{:}e_*^{-\pi i(2n+1)w}{:}_{-\tau}.
$$
Thus, we see the following:
\begin{thm}\label{Gammainvsin}
By altering the path of integration of l.h.s. by using the double-valued nature
of $\delta_*(w)$ so that it obtains the alternating periodicity $w\to w{+}1$, we have  
\begin{equation}\label{sinGamma}
2\int_{-\infty}^{\infty}\frac{e_*^{tw}}{1{+}e^t}dt
=
\pi(\sin_{*+}^{-1}\pi w +\sin_{*-}^{-1}\pi w).
\end{equation} 
\end{thm}
In general, it is dangerous to express such identity without showing 
the expression parameters. But, it should be permitted when the 
identity is obtained via several rotations of expression parameters. 

\bigskip
We often use rotations of path of integration together 
with rotations of expression parameters  
to observe the branching/periodic
behavior of the object. 

In the later chapter, we will show that elements 
which changes sign under the joint parallel 
displacement along a closed curve have a similar properties 
that are called {fermion} in physics. 

\bigskip
The notion of {\it parity} is an important notion in 
quantum physics. This is explained as the  
statistical characters of particles. However, such objects 
are treated in usual calculus. Therefore, one can ask 
as a mathematical question 

\begin{center}
``Is such a notion mathematically consistent ?'' 
\end{center}

We think this question must be proved mathematically. That means 
such a notion should be defined mathematically in 
ordinary calculus, just as the geometric notions of 
``length'',``area'' or ``volume''.

\subsection{Star-exponential functions of order $<2$}
\label{order<2}

Let $h(x)$ be a slowly increasing smooth function on ${\mathbb R}$ of 
growth order  
$\alpha$, $\alpha<2$, such as $\sqrt{1{+}x^2}$ or 
$\frac{x^3}{x^2+1}$. Formula   
\eqref{eq:Fouriertrsf} gives that if ${\rm{Re}}\tau>0$, 
then for every $s{\in}{\mathbb C}$, 
$e^{sh(x)}{:}\delta_*(x{-}w){:}_{\tau}$ 
is rapidly decreasing w.r.t. $x$. Hence, the integral 
\begin{equation}\label{starexpless2}
\int_{\mathbb R}e^{sh(x)} 
{:}\delta_*(x{-}w){:}_{\tau}dx
\end{equation}
is well-defined. 

\begin{thm}\label{Starexpless2}
If ${\rm{Re}}\tau>0$, 
$\int_{\mathbb R}e^{sh(x)}{:}\delta_*(x{-}w){:}_{\tau}dx$ satisfies the 
exponential law w.r.t. $s$. Moreover, this satisfies the 
evolution equation 
$$
\frac{d}{ds}f_s(w){=}{:}h_*(w){:}_\tau*_{\tau}f_s(w), \quad 
f_0(w)=1,
$$
where $h_*(w)=\int h(x)\delta_*(x{-}w)dx$. Hence, 
this gives a complex one parameter group. 
Thus, we denote 
this integral \eqref{starexpless2} by 
$e_*^{sh_*(w)}$, $s\in{\mathbb C}$.
\end{thm}

\noindent
{\bf Proof}\,\, 
Differentiating \eqref{starexpless2} by $s$, we have 
$$
\int_{\mathbb R}h(x)e^{sh(x)}\delta_*(x{-}w)dx. 
$$ 
Since $h_*(w)=\int h(x)\delta_*(x{-}w)dx$, we have 
$$
h_*(w){*}\int_{\mathbb R}e^{sh(x)}{:}\delta_*(x{-}w){:}_{\tau}dx=
\iint h(x')e^{sh(x)}{:}\delta_*(x'{-}w){*}\delta_*(x{-}w){:}_{\tau}dx'dx
$$
Since \eqref{sekidlt} gives 
$\delta_*(x'{-}w){*}\delta_*(x{-}w)=\delta(x{-}x')\delta_*(x{-}w)$ 
in the sense of distribution, this is 
$$
\int h(x)e^{sh(x)}{:}\delta_*(x{-}w){:}_{\tau}dx=
\partial_s\int_{\mathbb R}h(x)e^{sh(x)}\delta_*(x{-}w)dx
$$
It follows 
$$
\frac{d}{ds}e_*^{sh_*(w)}=h_*(w){*}e_*^{sh_*(w)}.
$$
The exponential law is verified directly by the computations 
as follows:
$$
e_*^{sh_*(w)}{*}e_*^{s'h_*(w)}=
\iint e^{sh(x){+}s'h(y)}\delta_*(x{-}w){*}\delta_*(y{-}w)dxdy
=\iint e^{sh(x){+}s'h(x)}\delta_*(x{-}w)dxdy.
$$
${}$\hfill $\Box$

In the definition above, the rotation of 
expression parameters cannot be 
considered in general. 
However, if $h(x)$ is a rational 
function such as $\frac{x^3}{(x{+}1+i)(x+1-i)}$, certain complex
rotations allowed together with rotations of expression 
parameters.
$$
\frac{1}{2\pi}\int_{\mathbb R}
e^{sh(e^{i\theta}x)}
{:}\delta_*(e^{i\theta}x-w){:}_{|\tau|e^{i\sigma}}
d(e^{i\theta}x) 
$$
is well-defined for $(\theta,\sigma)$ such that 
 $|-2\theta{+}\sigma|<\frac{\pi}{2}$,  
$e^{i\theta}(1\pm i)\not\in{\mathbb R}$, and this 
is independent of $\theta$ in this domain. 
Thus, the value of the integral has a discontinuous jump 
given by the residues at the singular points.

In such a case, one can consider the 
{\bf joint parallel displacement} along a curve in the 
expression parameter space by setting 
$\theta=\frac{\sigma}{2}$.

\subsection{Star-exponential function of $w_*^2$}\label{expw2}
As we have seen, the ${*}$-exponential function $e_*^{sh_*(w)}$ is 
very naive for the order of $h(x)$ is less than $2$. 
In this section, we treat the $*$-exponential function 
of the quadratic form $w_*^2$. 
Noting that 
${:}{w}_*^2{:}_{\tau}={w}^2{+}\frac{\tau}{2}$ in the 
$\tau$-expression, we now define 
the star-exponential function of ${w}_*^2$ 
by the real analytic solution of the evolution equation 
\begin{equation}\label{expdiffeq}
\frac{d}{dt}f_t={:}{w}_{*}^2{:}_{\tau}{*_{\tau}}f_t, 
\quad  f_0 = 1.
\end{equation} 
Precisely, this is 
$$
\frac{d}{dt}f_t=
\frac{\tau^2}{4}f''_t{+}{\tau}wf'_t
{+}({w}^2{+}\frac{\tau}{2})f_t,\quad f_0=1.
$$
To solve this, we set 
${:}f_t{:}_{\tau}=g(t)e^{h(t){w}^2}$,
by using the uniqueness of real analytic solutions.
Then, we have a system of the ordinary differential equations: 
\begin{equation}\label{system}
\left\{
\begin{aligned}
&\frac{d}{dt}h(t)=(1{+}\tau h(t))^2, \qquad\qquad h(0)=0\\
&\frac{d}{dt}g(t)=
\frac{1}{2}(\tau^2h(t){+}\tau)g(t),\qquad g(0)=1.  
\end{aligned}
\right. 
\end{equation}
The solution 
${:}e_*^{{t}{w}_*^2}{:}_{\tau}$ is given by   
\begin{equation}
 \label{eq:expquad}
{:}e_*^{{t}{w}_*^2}{:}_{\tau}=
\frac{1}{\sqrt{1{-}\tau t}}\,
e^{\frac{t}{1{-}\tau t}{w}^2}, 
\text{ for }\forall \tau, \,\, t\tau\not=1,\quad
{\text{(double-valued)}}.
\end{equation}
Note that no restriction to $\tau$.
$e_*^{tw_*^2}$ is obtained for all $\tau$.
This solution is obtained also via the intertwiner 
$I_0^{\tau}e^{tw^2}$. (See \eqref{2-to-2}.) 
The Weyl ordered expression ($\tau{=}0$-expression) gives   
${:}e_*^{t{w}_*^2}{:}_{0}{=}e^{t{w^2}}$ 
without singular point, and the $\tau$-expression is  
${:}e_*^{t{w}_*^2}{:}_{\tau}{=}
I_0^{\tau}(e^{t{w^2}})$, but note that these 
have singularity at $t\tau{=}1$. 

It is rather surprising that the solution has 
a branching singular point and hence, this does not form a 
complex one parameter group whenever $\tau\not=0$ is fixed.  
Moreover, the solution is double-valued w.r.t. the variable $t$. 

Since 
$$
\lim_{t\to\infty}\frac{\sqrt{t}}{\sqrt{1{-}t\tau}}e^{\frac{t}{1{-}t\tau}}=
\frac{1}{\sqrt{-\tau}}e^{-\frac{1}{\tau}w^2},
$$ 
we see that $\frac{1}{\sqrt{-\tau}}e^{-\frac{1}{\tau}w^2}$ is
invariant by $e_*^{s{w}_*^2}$, that is, 
$$
{:}e_*^{s{w}_*^2}{:}_{\tau}{*_{\tau}}\frac{1}{\sqrt{-\tau}}e^{-\frac{1}{\tau}w^2}
{=}\frac{1}{\sqrt{-\tau}}e^{-\frac{1}{\tau}w^2}.
$$

\bigskip 
Double-valued nature of $e_*^{t{w}_*^2}$ makes it difficult 
to select the domain where $e_*^{s{w}_*^2}+e_*^{t{w}_*^2}$ is 
a single valued function of two independent variables $(s,t)$. 

Moreover, it should be very careful to treat elements such as 
$$
\sinh_*sw_*^2=\frac{1}{2}(e_*^{s{w}_*^2}{-}e_*^{{-}s{w}_*^2}), \quad 
\cosh_*sw_*^2=\frac{1}{2}(e_*^{s{w}_*^2}{+}e_*^{{-}s{w}_*^2})
$$
for these are not double-valued, but triple valued functions of $s$ and
these two have no difference. Since 
$$
{:}\sinh_*sw_*^2{:}_{\tau}=
\frac{1}{2\sqrt{1{-}s\tau}}e^{\frac{s}{1{-}s\tau}w^2}{-}
\frac{1}{2\sqrt{1{+}s\tau}}e^{-\frac{s}{1{+}s\tau}w^2}
$$
there are two singular points $s=\pm\tau^{-1}$ and we can choose the
sign $\sqrt{1{-}s\tau}$, $\sqrt{1{+}s\tau}$ 
independently. Hence, ${:}\sinh_*0w_*^2{:}_{\tau}$ may be seen as 
$1, 0, {-}1$. Thus, we have to restrict the domain for $s$  to treat 
an element such as 
$$
f(s)=\sum_{n}{:}a_ne_*^{c_nsw_*^2}{:}_{\tau},\quad a_n, c_n\in{\mathbb C}
$$
as a single valued function.

In spite of such a difficulty, if a continuous curve $C$ does not hit
singular points, then ${:}e_*^{tw_*^2}{:}_{\tau}$ can be treated as a 
continuous function on $C$. Hence, the uniqueness of 
real analytic solution gives the exponential law 
$e_{*}^{sw_*^2}{*}e_{*}^{tw_*^2}=e_{*}^{(s+t)w_*^2}$:
$$
\frac{1}{\sqrt{1{-}\tau s}}\,e^{\frac{s}{1{-}\tau s}{w}^2}
{*_{\tau}}
\frac{1}{\sqrt{1{-}\tau t}}\,e^{\frac{t}{1{-}\tau t}{w}^2}
=
\frac{1}{\sqrt{1{-}\tau(s{+}t)}}\,e^{\frac{s{+}t}{1{-}\tau(s{+}t)}{w}^2}.
$$ 
Indeed, this is obtained by solving \eqref{system} with initial data 
$h(0)=\frac{t}{1{+}\tau t}$, $g(0)=\frac{1}{\sqrt{1{-}t\tau}}$ combined
with calculations such as 
$\sqrt{a}\sqrt{b}=\sqrt{ab}$, $\sqrt{a}/\sqrt{a}=\sqrt{1}=\pm 1$. 

Similarly, we have the exponential law  $e^{s}{*}e_{*}^{tw_*^2}=e_{*}^{s+tw_*^2}$
with the ordinary scalar exponential functions.

Recall that if a continuous curve $C$ does not hit
singular points, then ${:}e_*^{tw_*^2}{:}_{\tau}$ can be treated as a 
continuous function on $C$. Hence, one can treat the integral 
$\int_C{:}e_*^{tw_*^2}{:}_{\tau}dt$ without ambiguity.  
By this reason, it is better to 
define $\sinh_*sw_*^2$, $\cosh_*sw_*^2$ by the integral 
\begin{equation}\label{defsincos}
\sinh_*sw_*^2=\frac{1}{2}\int_{-s}^{s}w_*^2{*}e_*^{tw_*^2}dt,\quad 
\cosh_*sw_*^2=\frac{1}{2}\frac{d}{ds}\int_{-s}^{s}e_*^{tw_*^2}dt
\end{equation} 
via operations in the continuous calculus.

\bigskip
Formula \eqref{eq:expquad} is easily inverted by setting
$\frac{t}{1{-}t\tau}=a$ to obtain 
\begin{equation}\label{inverted}
e^{aw^2}=
{:}\frac{1}{\sqrt{1+a\tau}}\,\,
e_*^{\frac{a}{1{+}a\tau}w_*^2}{:}_{\tau}.
\end{equation} 
Thus, this makes calculations of $*_{\tau}$-product easy 
for exponential functions of quadratic forms by the 
exponential law:
\begin{equation}\label{easicalculation}
e^{aw^2}{*_{\tau}}e^{bw^2}=
\frac{1}{\sqrt{(1+a\tau)(1+b\tau)}}\,\,
{:}e_*^{(\frac{a}{1{+}a\tau}{+}\frac{b}{1{+}b\tau})w_*^2}{:}_\tau. 
=
\frac{1}{\sqrt{1-ab\tau^2}}
e^{\frac{a{+}b{+}2ab\tau}{1{-}ab\tau^2}w^2}
\end{equation}

\bigskip

So far, the expression parameter $\tau$ is fixed, but 
the double-valued nature of ${:}e_*^{tw_*^2}{:}_{\tau}$  
appears in the parallel translation along expression parameters. 

The next one may sound strange:
\begin{prop}
The identity $1$ is connected to $-1$ by a parallel 
translation along a closed curve of expression parameters. 
However, the identity $1$ here is not the absolute 
scalar but the identity element of the group 
$\{e_*^{tw_*^2}\}$.
\end{prop}
We can locally distinguish the $\pm$ sign, 
but we cannot globally.

\bigskip
\noindent 
{\bf Note for a change of generator}\,\,
Take a new generator, $\hat w$ to be $\hat w=aw$. Then, we see  
\begin{equation}\label{LagLag}
{:}e_*^{t{\hat w}_*^2}{:}_{a^2\tau}=
\frac{1}{\sqrt{1{-}ta^2\tau}}e^{\frac{t}{1{-}ta^2\tau}a^2w^2}=
{:}e_*^{ta^2w_*^2}{:}_{\tau}\quad c.f.\eqref{genchnge}.
\end{equation}
Set $t=-s\frac{1}{a^2}$ and $\tau=-1$ to obtain  
$$
{:}e_*^{t{\hat w}_*^2}{:}_{-a^2}=
\frac{1}{\sqrt{1{-}ta^2\tau}}e^{\frac{t}{1{-}ta^2\tau}a^2w^2}=
{:}e_*^{-sw_*^2}{:}_{-1}\quad c.f.\eqref{genchnge}.
$$
Hence, every ${:}e_*^{tw_*^2}{:}_{\tau}$ is obtained from 
the generating function of the Laguerre polynomials.

\subsubsection{The generating function  of Laguerre polynomials}

The generating function of Laguerre polynomials 
$L_n^{(\alpha)}(x)$ is given as follows:
$$
\frac{1}{(1{-}t)^{\alpha+1}}e^{-\frac{t}{1{-}t}x}= 
\sum_{n\geq 0}L_n^{(\alpha)}(x)t^n, 
\quad (|t|<1).
$$
Differentiating both sides by $x$, we have also the relations 
\begin{equation}\label{Laguerre}
\frac{d}{dx}L_n^{(\alpha)}(x)=-L_{n-1}^{(\alpha{+}1)}(x).
\end{equation}

If $\alpha{=}-\frac{1}{2}$, 
this is the $\tau=-1$ expression of $e_*^{-tw_*^2}$, i.e.
$$
{:}e_*^{-tw_*^2}{:}_{{-}1}
{=}\frac{1}{(1{-}t)^{\frac{1}{2}}}e^{{-}\frac{t}{1{-}t}w^2}
{=}\sum_{n\geq 0}L_n^{(-\frac{1}{2})}(w^2)t^n.
$$
Setting $x=w^2$, we see 
$$
\frac{d}{dx}{:}e_*^{-tw_*^2}{:}_{{-}1}
=\frac{1}{(1{-}t)^{\frac{1}{2}}}
\frac{d}{dx}e^{{-}\frac{t}{1{-}t}x} 
{=}{-}\sum_{n\geq 0}L_n^{(\frac{1}{2})}(x)t^{n+1}.
$$

Minding these, we define $*$-Laguerre polynomials 
$L_n(w^2,\tau)={:}L_n(w^2,*){:}_{\tau}$ by 

$$
e_*^{tw_*^2}=
\sum_n L_n^{(-\frac{1}{2})}(w^2,*)\frac{1}{n!}t^n,\quad 
L_n^{(-\frac{1}{2})}(w^2,\tau)=
\frac{d^n}{dt^n}\Big|_{t=0}
\frac{1}{(1{-}t\tau)^{\frac{1}{2}}}
e^{\frac{1}{\tau(1{-}t\tau)}w^2}e^{-\frac{1}{\tau}w^2}. 
$$
As $t=0$ is a regular point, these are well-defined, and the
exponential law gives 
$$
L_n^{(-\frac{1}{2})}(w^2,*)=\sum_{k{+}\ell=n}L_k^{(-\frac{1}{2})}(w^2,*)
{*}L_\ell^{(-\frac{1}{2})}(w^2,*).
$$

Note that setting $x=w^2$, 
$$
\frac{d}{dt}\frac{x^{\alpha{-}1}}{(1{-}t\tau)^{\alpha}}e^{\frac{1}{\tau(1-t\tau)}}
=
\frac{d}{dx}\frac{1}{\tau}
\frac{x^{\alpha}}{(1{-}t\tau)^{\alpha{+}1}}e^{\frac{1}{\tau(1-t\tau)}}.
$$
Using this, we see that 
$$
\frac{d^n}{dt^n}\Big|_{t=0}
\frac{1}{(1{-}t\tau)^{\frac{1}{2}}}
e^{\frac{1}{\tau(1{-}t\tau)}w^2}e^{-\frac{1}{\tau}w^2}
=\Big(\tau^{-n}\frac{d^n}{dx^n}(x^{\frac{1}{2}{+}n}e^{\frac{1}{\tau}x})\Big)
x^{-\frac{1}{2}}e^{-\frac{1}{\tau}x}.
$$
It follows that setting $x=w^2$
$$
L_n^{(-\frac{1}{2})}(w^2,\tau)=
\frac{1}{n!}\Big(\tau^{-n}\frac{d^n}{dx^n}(x^{\frac{1}{2}{+}n}e^{\frac{1}{\tau}x}\Big)
x^{-\frac{1}{2}}e^{-\frac{1}{\tau}x}
$$

As in the case of Hermite polynomials, this formula is used to 
to obtain the orthogonality property of $L_n^{(-\frac{1}{2})}(w^2,\tau)$. 

Assuming that ${\rm{Re}}\tau<0$ and restricting $x=w^2$ to the real axis,
we want to show that 
$$
\int_{\mathbb R}x^{\frac{1}{2}}e^{\frac{1}{\tau}x}
L_n^{(-\frac{1}{2})}(x,\tau)L_m^{(-\frac{1}{2})}(x,\tau)dx=\delta_{n,m}.
$$
First remark that $L_n(x,\tau)$ is a polynomial of degree $n$.
$$
\int_{\mathbb R}x^{\frac{1}{2}}e^{\frac{1}{\tau}x}L_n^{(-\frac{1}{2})}(x,\tau)L_m^{(-\frac{1}{2})}(x,\tau)dx=
\int_{\mathbb R}\frac{1}{\tau^n}\frac{1}{n!}\Big(\frac{d^n}{dx^n}
x^{\frac{1}{2}{+}n}e^{\frac{1}{\tau}{x}}\Big)L_m^{(-\frac{1}{2})}(x,\tau)dx.
$$
If $n\not= m$, one may assume that $n>m$ without loss of generality. 
Hence, this vanishes by integration by parts $n$ times.

For the case $n=m$, recalling $L_n^{(-\frac{1}{2})}(x,\tau)$ is a
polynomial of degree $n$, and 
taking $\frac{d^n}{dx^n}$ of both sides of the equality in the next line: 
$$
L_n^{(-\frac{1}{2})}(x,\tau)=
\frac{1}{n!}\frac{d^n}{dt^n}\Big|_{t=0}
\frac{1}{(1{-}t\tau)^{\frac{1}{2}}}
e^{\frac{1}{\tau(1{-}t\tau)}x}e^{-\frac{1}{\tau}x},
$$
we have 
$$
\begin{aligned}
\frac{d^n}{dx^n}L_n^{(-\frac{1}{2})}(x,\tau)
=&\frac{1}{n!}\frac{d^n}{dt^n}\Big|_{t=0}\frac{d^n}{dx^n}
\frac{1}{(1{-}t\tau)^{\frac{1}{2}}}e^{\frac{t}{1{-}t\tau}x}\\
=&\frac{1}{n!}\frac{d^n}{dt^n}\Big|_{t=0}
\frac{t^n}{(1{-}t\tau)^{\frac{1}{2}{+}n}}e^{\frac{t}{1{-}t\tau}x}.
\end{aligned}
$$
But the last term does not contain $x$ for this must be degree $0$. 
Hence, 
$$
\frac{d^n}{dx^n}L_n^{(-\frac{1}{2})}(x,\tau)
=\frac{1}{n!}\frac{d^n}{dt^n}\Big|_{t=0}
\frac{t^n}{(1{-}t\tau)^{\frac{1}{2}{+}n}}=1.
$$

\subsubsection{Intertwiners are 2-to-2 mappings}
 
The intertwiner $I_{\tau}^{\tau'}$ is defined by 
$e^{\frac{1}{4}(\tau'{-}\tau)\partial^2_w}$. 
The case of exponential functions of quadratic forms
is treated by solving the evolution equation 
$$
\frac{d}{dt}f_t(w)=\partial_w^2 f(w),\quad f_0(w)=ce^{aw^2}.
$$
Setting $f_t=g(t)e^{q(t)w^2}$, this equation becomes 
$$
\left\{
\begin{matrix}
\frac{d}{dt}q(t)= 4q(t)^2  & q(0)=a \\
{}                         &{}\\
\frac{d}{dt}g(t)= 2g(t)q(t)& g(0)=c \\
\end{matrix}
\right.
$$
Solving this gives 
$g(t)e^{q(t)w^2}=
\frac{c}{\sqrt{1-4ta}}e^{\frac{a}{1-4ta}w^2}$.
Plugging in $t=\frac{1}{4}(\tau'-\tau)$, we obtain 
$$
I_{\tau}^{\tau'}(ce^{aw^2})=
\frac{c}{\sqrt{1{-}(\tau'{-}\tau)a}}
e^{\frac{a}{1{-}(\tau'{-}\tau)a}w^2}.
$$ 
Note that this also covers the case 
for intertwiners for 
$e_*^{\frac{t}{i\h}\langle{\pmb a},{\pmb u}\rangle}$. 

\medskip
To reveal its double-valued nature, we rewrite this as follows:
\begin{equation}\label{2-to-2}
I_{\tau}^{\tau'}
(\frac{c}{\sqrt{1{-}\tau t}}e^{\frac{t}{1-\tau t}w^2})
=\frac{c}{\sqrt{1{-}\tau' t}}e^{\frac{t}{1-\tau' t}w^2}.
\end{equation} 
Since the branching singular point of 
the double-valued section of  
the source space moves by the intertwiners, $I_{\tau}^{\tau'}$ 
must be viewed as a $2$-to-$2$ mapping.

\medskip
To explain the detail, we construct two sheets with slit from
$\tau^{-1}$ to $\infty$, and denote points by $(t;+)_{\tau}$ or $(t;-)_{\tau}$.
$I_{\tau}^{\tau'}$ has the property that 
$I_{\tau}^{\tau'}((t;\pm)_{\tau})=(t;\pm)_{\tau'}$ as a set-to-set
mapping, and one may define this locally a 1-to-1 mapping. 
Note that 
$$
I_{\tau''}^{\tau}I_{\tau'}^{\tau''}I_{\tau}^{\tau'}((t,\pm)_{\tau})=(t,\pm)_{\tau},
$$ 
but this is neither the identity nor $-1$. This depends on $t$ discontinuously.

\bigskip
On the other hand, we want to retain the feature of complex one parameter group. 
For that purpose, we have to set ${:}e_*^{0w_*^2}{:}_{\tau}=1$, the multiplicative unit 
for every expression. The problem is caused by another sheet, for we have to 
distinguish $1$ and $-1$.

It is important to recognize that 
there is no effective theory to understand such a vague 
system. This is something like an {\it air pocket} of the 
theory of point set topology. This is not a difficult 
object, but an object that can be treated case by cases 
to avoid its ambiguous character. However, the important 
thing is that such phenomena occur very often in 
the stage of applying calculus. 
As it will be seen in the next section,  
this system forms an object which may be viewed 
as a {\it double covering group} of ${\mathbb C}$. 
Well, this is absurd since ${\mathbb C}$ is simply 
connected !

\noindent
{\bf Note}\,\,A toy model of such a strange object is given by
considering Hopf fibering 
$S^3=\coprod_{x\in S^2}S_x^1$ where $S_x^1$ is the fiber at 
$x\in S^2$. Let $\tilde S_x^1$ be the double covering 
group of each fiber. Then, consider the disjoint sum 
$\coprod_{x\in S^2}\tilde S_x^1$.

\subsubsection{Another definition of $*$-inverses of $a^2{+}w_*^2$}

Since $a^2{+}w_*^2=(a{+}iw){*}(a{-}iw)$, its  
inverse may be defined by  
$$
(a{+}iw)_{*+}^{-1}{*}(a{-}iw)_{*-}^{-1}=\frac{1}{2a}((a{+}iw)_{+}^{-1}{+}(a{-}iw)_{-})^{-1},\quad 
{\rm{Re}}\,\tau>0.
$$
Hence, we have for $a\not=0$, 
$$
\begin{aligned}
\frac{1}{2a}((a{+}iw)_{+}^{-1}{+}(a{-}iw)_{-})^{-1}&=
\int_{-\infty}^0
\frac{1}{2a}(e_*^{t(a{+}iw)}{+}e_*^{t(a{-}iw)})dt\\
&=
\int_{-\infty}^0\frac{1}{a}e^{at}\cos_*(tw)dt,\quad {\rm{Re}}\,\tau>0.
\end{aligned}
$$

On the other hand, using 
$$
{:}e_*^{tw^2}{:}_{\tau}=\frac{1}{\sqrt{1{-}t\tau}}e^{\frac{t}{1{-}t\tau}w^2},\quad
\tau\in{\mathbb C}.
$$
 and the exponential law, we define another $*$-inverse of $a^2{+}w_*^2$, for $a^2>0$  by 
$$
(a^2{+}w_*^2)^{-1}=\int^{0}_{-\infty}e_*^{t(a^2{+}w_*^2)}dt,\quad \tau\not=0.
$$     
In precise,   
$$
\int^{0}_{-\infty}{:}e_*^{t(a^2{+}w_*^2)}{:}_{\tau}dt=
\int^{0}_{-\infty}\frac{e^{a^2t}}{\sqrt{1{-}t\tau}}e^{\frac{t}{1{-}t\tau}w^2}dt,
\quad a^2>0.
$$  
The difference of these two inverses satisfies 
$(a^2{+}w_*^2){*}f_*(w)=0$, hence it is given by a linear combination
of $\delta_*(w{+}ia)$ and $\delta_*(w{-}ia)$.

\subsection{Star-exponential functions of higher order}
\label{DefbyFT}

Note that a $*$-monomial $e^{im\theta}w_*^m$ may be defined by  
$e^{im\theta}w_*^m=\int e^{im\theta}x^m\delta_*(x{-}w)dx$, but noting that 
$e^{im\theta}w_*^m=(e^{i\theta}w)_*^m$, this may be defined as 
$\int x^m\delta_*(x{-}e^{i\theta}w)dx$. 
\begin{lem}\label{nontrividentity}
If both $f(x)$ and $f(e^{i\theta}x)$ are tempered distribution, then
it holds the equality
$$
\int f(x)\delta_*(x{-}e^{i\theta}w)dx=\int f(e^{i\theta}x)\delta_*(x{-}w)dx 
$$
In particular, it holds the identity 
$$
\int f(x)\delta_*(x{-}e^{i\theta}w)dx=\int f({-}x)\delta_*(x{+}e^{i\theta}w)dx 
$$
\end{lem}

\medskip
Consider the evolution equation 
$$
\frac{d}{dt}f_t(w)=iw_*^{\ell}{*}f_t(w),\quad f_0(w)=1
$$
for star-exponential functions of higher order $\ell\geq 3$. In
precise, this is considered under suitable $\tau$-expressions:  
$$
\frac{d}{dt}{:}f_t(w){:}_{\tau}=i{:}w_*^{\ell}{:}_{\tau}{*_\tau}{:}f_t(w){:}_{\tau},
\quad {:}f_0(w){:}_{\tau}=1.
$$
This is a partial differential equation of order $\ell\geq 3$, but 
 Proposition\,\ref{convzero} shows that there is no real analytic
 solution, if the expression parameter $\tau\not=0$.

However, as $e^{isx^{\ell}}$ is a slowly increasing function of $x\in{\mathbb R}$
for every real number $s$ and for every positive integer $\ell$, 
the formula \eqref{startemparete} gives that 
$\int_{\mathbb R}e^{isx^{\ell}}{:}\delta_*(x{-}w){:}_{\tau}dx$ 
for ${\rm{Re}}\,\tau>0$ is a one parameter group which may be 
denoted by $e_*^{isw_*^{\ell}}$.

As $e^{ise^{i\theta}x^{\ell}}$ is never slowly increasing if
$e^{i\theta}\not\in{\mathbb R}$, the parameter $s$ of 
$e_*^{isw_*^{\ell}}=\int_{\mathbb R}e^{isx^{\ell}}{:}\delta_*(x{-}w){:}_{\tau}dx$
cannot be complexified.  
However, note that  
\begin{equation}\label{higherexp}
\int_{\mathbb R}e^{isx^{\ell}}
\delta_*(x{-}e^{i\frac{\theta}{\ell}}w)dx =
e_*^{is(e^{i\frac{\theta}{\ell}}w)_*^{\ell}}.
\end{equation}
If r.h.s. is equal to $e_*^{ise^{i\theta}w_*^{\ell}}$, then we can 
complexify the parameter $s$, but in what follows we see this is 
not true. Note again that 
$$
{:}\delta_*(x{-}e^{\frac{i}{\ell}(\theta{+}2\pi)}w){:}_{\tau}\not=
{:}\delta_*(x{-}e^{\frac{i}{\ell}\theta}w){:}_{\tau}.
$$
We define 
\begin{equation}\label{eq:niceFdef}
I_{\ell}(s,\theta,\tau)=\int e^{isx^{\ell}}{:}\delta_*(x{-}e^{\frac{i}{\ell}\theta}w){:}_{\tau}dx=
\int_{\mathbb R}{\mathfrak F}^{-1}(e^{isx^{\ell}})(t){:}e_*^{-ite^{\frac{i}{\ell}\theta}w}{:}_{\tau}\dbar t,
 \quad 
{\rm{Re}}\,e^{2\frac{i}{\ell}\theta}\tau>0. 
\end{equation}
It is useful to keep the equality 
$I_{\tau}^{\tau'}I_{\ell}(s,\theta,\tau)=I_{\ell}(s,\theta,\tau')$
in mind. 

In this section we show that $I_{\ell}(s,\theta,\tau)$ is  
defined on some sector in the universal covering space 
${\widetilde{\mathbb C}}_*$ of 
${\mathbb C}{\setminus}\{0\}$ and it 
behaves as if it were an $\ell$-{\it covering group} of 
the $*$-exponential function $e_*^{ise^{i\theta}w_*^{\ell}}$.

\medskip
As  ${\rm{Re}}\,e^{2\frac{i}{\ell}\theta}\tau>0$, 
$e^{-\frac{1}{4}e^{2\frac{i}{\ell}\theta}\tau t^2}$ is 
a rapidly decreasing in $t$, and so also 
$e^{ite^{\frac{i}{\ell}\theta}w}
e^{-\frac{1}{4}e^{2\frac{i}{\ell}\theta}\tau t^2}$ on 
every compact domain of $w$. 
 Thus, this is well-defined for 
every $s\in{\mathbb R}$. 
It is easy to see that $I_{\ell}(0,\theta,\tau)=1.$

Integration by parts twice gives that 
\begin{equation}\label{intpart} 
\partial_{s}I_{\ell}(s,\theta,\tau){=}
ie^{i\theta}w_*^{\ell}{*}_{\tau}I_{\ell}(s,\theta,\tau), \quad 
\partial_{\theta}I_{\ell}(s,\theta,\tau){=}
(is)ie^{i\theta}w_*^{\ell}{*}_{\tau}I_{\ell}(s,\theta,\tau)
{=}is\partial_{s}I_{\ell}(s,\theta,\tau).
\end{equation} 
\eqref{intpart} gives also that 
$$
\frac{d^k}{d(e^{i\theta}s)^k}
I_{\ell}(s,\theta,\tau)\big|_{s=0}=
e^{-ik\theta}\partial^k_{s}I_{\ell}(s,\theta,\tau)\big|_{s=0}= 
{:}i^kw_*^{k\ell}{:}_{\tau}.
$$
Thus, the Taylor series of 
$I_{\ell}(s,\theta,\tau)$ at $s=0$ is 
the diverging series 
$\sum_k \frac{(ise^{i\theta})^k}{k!}
{:}w_*^{k{\ell}}{:}_\tau$ 
for $\ell\geq 3$ and $\tau\not=0$. 

\begin{lem}\label{polar}
If a complex valued smooth function 
$f(s,\theta)$ satisfies the partial differential equation 
$$
is\partial_sf(s,\theta)=\partial_{\theta}f(s,\theta).
$$
Then, $f(s,\theta)$ is a holomorphic function on the 
open subset   
$$
\{(s,\theta); s{\in}{\mathbb R}, \,\,
{\rm{Re}}\,e^{2\frac{i}{\ell}\theta}\tau>0\}
$$
of the universal covering space of $\mathbb C{\setminus}\{0\}$. 
\end{lem}

\noindent
{\bf Proof}\,\,Rewrite the Cauchy-Riemann equation by  
the polar coordinate system $se^{i\theta}$. Note that  
$$
\partial_x=\cos\theta \partial_s-
    \frac{1}{s}\sin\theta\partial_{\theta}, \quad 
\partial_y=\sin\theta \partial_s+
    \frac{1}{s}\cos\theta\partial_{\theta}.
$$  
The condition 
$is\partial_sf(s,\theta)=\partial_{\theta}f(s,\theta)$ 
shows that $f=u(s,\theta)+iv(s,\theta)$ 
satisfies the Cauchy-Riemann equation 
on the space $s\not=0$. 

Suppose $\tau=\tau_0e^{i\sigma}, \tau_0{>}0$. Then the 
defining domain is 
$|\frac{2}{\ell}\theta{+}\sigma|{<}\frac{\pi}{2}$. It is 
precisely
\begin{equation}\label{domainexp}
D_{\sigma}=\{(s,\theta,\sigma); s{\in}{\mathbb R}, 
-\frac{\ell}{4}\pi-\frac{\ell\sigma}{2}<
\theta<\frac{\ell}{4}\pi-\frac{\ell\sigma}{2}\}
\end{equation}
in the universal covering space of $\mathbb C{\setminus}\{0\}$
depending on $\sigma$. 
\hfill$\Box$

\subsubsection{$\ell$-covering property }

Consider the defining domain \eqref{domainexp} of
$I_{\ell}(s,\theta,\tau)$. 
If $\ell$ is big enough, say $\ell>8$, then 
$D_{\sigma}$ for any fixed $\sigma$ contains an interval 
$[\theta,\theta{+}2\pi]$. Hence, 
$I_{\ell}(s,\theta,\tau)$ and $I_{\ell}(s,\theta{+}2\pi,\tau)$ 
are defined under the same expression parameter.
If these are equal, then 
$I_{\ell}(s,\theta,\tau)={:}e_*^{se^{i\theta}w^{\ell}_*}{:}_{\tau}$ 
is well-defined on $se^{i\theta}\in{\mathbb C}{\setminus}\{0\}$, and 
the origin is a removable singular point. 
Since this against  Proposition\,\ref{convzero}, we have that 
$I_{\ell}(s,\theta,\tau)\not=I_{\ell}(s,\theta{+}2\pi,\tau)$. 

Using the second identity of Lemma\,\ref{nontrividentity}
one can make the same argument in case that $D_{\sigma}$ 
contains an interval $[\theta,\theta{+}\pi]$, but 
if we consider intertwiners together with the joint parallel
translations, we can conclude that 
$$
I_{\ell}(s,\theta{+}2\pi,\tau')\not=I_{\tau}^{\tau'}I_{\ell}(s,\theta,\tau).
$$
On the other hand, as 
$I_{\ell}(s,\theta{+}2\pi\ell,\tau)=\int e^{isx^{\ell}}
{:}\delta_*(x{-}e_*^{i\frac{\theta{+}2\pi\ell}{\ell}w}){:}_{\tau}dx=
\int e^{isx^{\ell}}{:}\delta_*(x{-}e_*^{i\frac{\theta}{\ell}w}){:}_{\tau}dx$ we have 
$$
I_{\ell}(s,\theta{+}2\pi\ell,\tau)=I_{\ell}(s,\theta,\tau).
$$
Thus, we conclude the following: 
\begin{prop}\label{ellcovering}
What is defined by $I_{\ell}(s,\theta,\tau)$ is not
$e_*^{ise^{i\theta}w_*^{\ell}}$ but its branched $\ell$-covering.
\end{prop}

On the other hand, it holds for every $k$ 
$$
\begin{aligned}
\frac{d}{ds}I_{\ell}(s,\theta{+}2\pi k,\tau)&=
\int ix^{\ell}e^{isx^{\ell}}
{:}\delta_*(x{-}e^{i\frac{\theta{+}2\pi k}{\ell}}w){:}_{\tau}dx\\
&=
\int ix^{\ell}
{:}\delta_*(x{-}e^{i\frac{\theta{+}2\pi k}{\ell}}w){:}_{\tau}dx
{*}I_{\ell}(s,\theta{+}2\pi k,\tau). 
\end{aligned}
$$

We saw
$I_{\ell}(s,\theta{+}2\pi k,\tau)\not=I_{\ell}(s,\theta,\tau)$, but 
as $w_*^{\ell}$ is a polynomial, we see  
$$
\int ix^{\ell}
{:}\delta_*(x{-}e^{i\frac{\theta{+}2\pi k}{\ell}}w){:}_{\tau}dx=
i(e^{i\frac{1}{\ell}\theta}w_*)^{\ell}=ie^{i\theta}w_*^{\ell}
$$
It follows for every $k$
$$
\frac{d}{ds}I_{\ell}(s,\theta{+}2\pi k,\tau)= 
{:}ie^{i\theta}w_*^{\ell}{:}_{\tau}{*_{\tau}}I_{\ell}(s,\theta{+}2\pi
k,\tau),\quad I_{\ell}(0,\theta{+}2\pi k,\tau)=1.
$$
Thus, we have the following:

\begin{prop}\label{failunicty}
The uniqueness fails in the evolution equation 
$$
\frac{d}{dt}f_t(w)=iw_*^{\ell}{*}f_t(w),\,\,f_0(w)=1.
$$ 
There are $\ell$-different solutions, which are not real analytic at $t=0$. 
\end{prop}

\noindent
{\bf Product structure}\,\,\,
As
$\delta_*(x{-}e_*^{i\frac{\theta}{\ell}w}){*}\delta_*(x'{-}e_*^{i\frac{\theta}{\ell}w})
=\delta(x{-}x')\delta_*(x'{-}e_*^{i\frac{\theta}{\ell}w})$, we see 
$$
\int
e^{isx^{\ell}}\delta_*(x{-}e_*^{i\frac{\theta}{\ell}w})dx
\,{*}\!
\int
e^{is'{x'}^{\ell}}\delta_*(x'{-}e_*^{i\frac{\theta}{\ell}w}){:}_{\tau}dx'
{=}
\int
e^{i(s{+}s')x^{\ell}}\delta_*(x{-}e_*^{i\frac{\theta}{\ell}w})dx.
$$
However, if $\theta\not=\theta'$, then  the product 
$$
\int
e^{isx^{\ell}}\delta_*(x{-}e_*^{i\frac{\theta}{\ell}w})dx
\,{*}\!\int
e^{is'{x'}^{\ell}}\delta_*(x'{-}e_*^{i\frac{\theta'}{\ell}w}){:}_{\tau}dx'
$$
is not defined. 
It seems that there is no branch point other than the origin $s=0$.

\subsubsection{Star exponential functions of $(\alpha{+}w)_{*+}^{-1}$}
As we have seen already, the 
$*$-exponential function ${:}e_*^{isw}{:}_{\tau}$ has high regularity
w.r.t. $s\in{\mathbb C}$ when the expression parameters are 
restricted in a half-space, e.g. ${\rm{Re}}\,\tau>0$, 
and the $*$-inverse function $(\alpha{+}w)_{*+}^{-1}$ has high
regularity. Thus it seems natural to think that  
\begin{equation}
e_*^{s(\alpha{+}w)_{*+}^{-1}}=\sum\frac{s^{n}}{n!}\frac{(-1)^{n{-}1}}{(n{-}1)!}
\frac{d^{n-1}}{d\alpha^{n-1}}(\alpha{+}w)_{*+}^{-1}=
\sum_n\int_{-\infty}^0\frac{s^n({-}it)^{n{-}1}}{n!(n{-}1)!}e_{*}^{it(\alpha{+}w)}dt
\end{equation}
is defined with high regularity. 

This may be defined by the evolution equation
$\frac{d}{ds}f_s=((\alpha{+}w)_{*+}^{-1}){*}f_s$, $f_0=1$.
It is natural to think that this is included in the solution of 
\begin{equation}\label{invevol}
(\alpha{+}w){*}\frac{d}{ds}f_s(w)=f_s(w), \quad f_0(w)=1.
\end{equation}
But this equation ignores the $\pm$ ambiguities of the inverse. 

At first, solving
$$
(\alpha{+}w){*_{\tau}}g_{\lambda}(w)=
(\alpha{+}w)g_{\lambda}(w){+}\frac{\tau}{2}\partial_wg(w)
= \lambda g_{\lambda}(w),\quad \lambda{\in}{\mathbb C},
$$
we see that
$g_{\lambda}(w)=c{:}\delta_*(w{+}\alpha{-}\lambda){:}_{\tau}$,
$c\in{\mathbb C}$. 

Moreover, if ${\rm{Im}}\,\lambda<0$, then 
$$
\begin{aligned}
(w{+}\alpha)_{*+}^{-1}{*}\delta_*(w{+}\alpha{-}\lambda)=&
i\int_{-\infty}^0\!\! ds
\int_{-\infty}^{\infty}\!\! dt\,\, e_*^{is(w{+}\alpha)}e_*^{it(w{+}\alpha)}e^{-it\lambda}\\
=&
\int_{-\infty}^{\infty}\!\!\! e_*^{i\sigma(w{+}\alpha-\lambda)}d\sigma\,\, 
i\int_{-\infty}^0 \!\!\!e^{i\lambda s}ds=
\frac{1}{\lambda}\delta_*(w{+}\alpha{-}\lambda).
\end{aligned}
$$
It follows that 
${:}\delta_*(w{+}\alpha{-}\lambda){:}_{\tau}
e^{\frac{1}{\lambda}s}$
is a real analytic solution of
$\frac{d}{ds}h_s(w)=(w{+}\alpha)_{*+}^{-1}{*}h_s(w)$ providing
$\lambda\not=0$. Hence, it may be written as 
$$
{:}e_{*}^{s(w{+}\alpha)_{*+}^{-1}}{*}\delta_*(w{+}\alpha{-}\lambda){:}_{\tau}
={:}\delta_*(w{+}\alpha{-}\lambda){:}_{\tau}e^{\frac{1}{\lambda}s},\quad
{\rm{Re}}\,\tau>0.
$$
But this does not imply the existence of $e_{*}^{s(w{+}\alpha)_{*+}^{-1}}$.

To adjust the initial condition, we set $\lambda=\alpha{+}x{-}i\eta$,
$\eta>0$, and 
$$
f_s(w)=\int_{-\infty}^{\infty}e^{s\frac{1}{\alpha{+}x{-}i\eta}}{:}\delta_*(w{-}x{+}i\eta){:}_{\tau}
dx. 
$$
This is well-defined and independent of $\eta$ by Cauchy's integration theorem whenever $\eta>0$. 

This is the solution of \eqref{invevol} satisfying 
$f_0(w)=1$ by $\int_{-\infty}^{\infty}{:}\delta_*(w{-}x{+}i\eta){:}_{\tau}dx=1$
 and this gives also the solution of 
$$
\frac{d}{ds}\tilde{f}_s(w)=(w{+}\alpha)_{*+}^{-1}{*}\tilde{f}_s(w),\quad \tilde{f}_0(w)=1.
$$
Hence, 
$$
{:}e_*^{s(w{+}\alpha)_{*+}^{-1}}{:}_{\tau}=
\int_{-\infty}^{\infty}e^{s\frac{1}{\alpha{+}x{-}i\eta}}{:}\delta_*(w{-}x{+}i\eta){:}_{\tau}
dx.  \,\,\, s\in{\mathbb C}
$$
is a complex one parameter group. 

As  ${\rm{Im}}\,\lambda<0$, we see 
$\lim_{t\to-\infty}e^{-\frac{1}{\lambda}it}=0$, hence 
$$
\lim_{t\to\infty}e_*^{-it(w{+}\alpha)_{*+}^{-1}}=0,
\quad{\text{and}}\quad  
i\int_{-\infty}^0 e_*^{it(w{+}\alpha)_{*+}^{-1}}dt=w{+}\alpha=\Big((w{+}\alpha)_{*+}^{-1}\Big)_{*+}^{-1}.
$$


\begin{thebibliography}{OMMY2}
 
\bibitem[AAR]{AAR}{G. Andrews, R. Askey, R. Roy},\,\,
\newblock{\sc Special functions},
\newblock{Encyclopedia Math, Appl. 71, Cambridge, 2000.}


\bibitem[BF]{BF}{F.Bayen,\,M,Flato,\,C.Fronsdal,\,A.Lichnerowicz,\,D.Sternheimer,\,}
{\it Deformation theory and quantization I, II}, Ann. Phys. 111, (1977),
61-151.  


\bibitem[GS]{GS}{I.M.Gel'fand,\,G.E.Shilov,\,\,}
{\sc Generalized Functions, 2}, Acad. Press, 1968. 
 
\bibitem[M]{Mm}{M.Morimoto,\,\,}{\sc An introduction to Sato's hyperfunctions},
  AMS Trans. Mono.129, 1993.  

\bibitem[O]{om6}{H. Omori},\,\,{\it Toward geometric quantum theory}, in 
From Geometry to Quantum Mechanics. Prog. Math. 252,
Birkh{\"a}user, (2007), 213-251.    


{\bibitem[OMMY]{ommy}H. Omori, Y. Maeda, N. Miyazaki, A. Yoshioka}, 
\newblock{\it Deformation quantization of Fr\'echet-Poisson algebras, 
--{Convergence of the Moyal product--}},  
in  Conf\'erence Mosh\'e Flato 1999, Quantizations, Deformations, 
and Symmetries, Vol II,  Math. Phys. Studies 22, 
Kluwer Academic Press, (2000), 233-246.


\bibitem[OMMY2]{OMMY6}
{H.Omori, Y.Maeda, N.Miyazaki and A.Yoshioka :}
\newblock{\it Star exponential functions as two-valued elements}, 
  \newblock{in The breadth of symplectic and Poisson geometry}, 
  \newblock{Progress in Math. 232, Birkh{\"auser},(2004)}, 483-492. 




\bibitem[R]{Ri} 
{F. S. Ritt}
{\it On derivative of functions at a point.} 
 Ann. Math. 18, (1916) 18-23.
 

\end{thebibliography}
\end{document}